%% file: main.tex
\documentclass[fleqn,10pt]{wlscirep}
\usepackage[utf8]{inputenc}
\usepackage[T1]{fontenc}
\usepackage[symbol]{footmisc}
\usepackage{amsmath}
\usepackage{changepage}
\usepackage{color}
\usepackage{fancyhdr}
\usepackage{graphicx}
\usepackage{hyperref} 
\usepackage{rotating}
\usepackage{xcolor}
\usepackage[sort, numbers]{natbib}

\title{Incorporating Connectivity among Internet Search Data for Enhanced Influenza-like Illness Tracking}

\author[1,*]{Shaoyang Ning}
\author[1,+]{Ahmed Hussain}
\author[2]{Qing Wang}
\affil[1]{Department of Mathematics and Statistics, Williams College, Williamstown, MA, 01267, USA}
\affil[2]{Department of Mathematics, Wellesley College, Wellesley, MA, 02481, USA}

\affil[*]{sn9@williams.edu}

\affil[+]{This author contributed to the project as an undergraduate research assistant at Williams College}

\begin{abstract}
Big data collected from the Internet possess great potential to reveal the ever-changing trends in society. In particular, accurate infectious disease tracking with Internet data has grown in popularity, providing invaluable information for public health decision makers and the general public. However, much of the complex connectivity among the Internet search data is not effectively addressed among existing disease tracking frameworks. To this end, we propose ARGO-C (Augmented Regression with Clustered GOogle data), an integrative, statistically principled approach that incorporates the clustering structure of Internet search data to enhance the accuracy and interpretability of disease tracking. Focusing on multi-resolution \%ILI (influenza-like illness) tracking, we demonstrate the improved performance and robustness of ARGO-C over benchmark methods at various geographical resolutions. We also highlight the adaptability of ARGO-C to track various diseases in addition to influenza, and to track other social or economic trends.
\end{abstract}
\begin{document}

\flushbottom
\maketitle
%
%
\thispagestyle{empty}


\section*{Introduction}
Big data collected from the Internet, recording billions of people's digital footprints, possess great potential to reveal the ever-changing trends in society. A growing number of attempts have been made to harness the potential of Internet data to address issues in a wide range of fields, including public health \cite{polgreen2008using, ginsberg2009detecting, althouse2011prediction, chan2011using,
murdoch2013inevitable, lee2013real, khoury2014big, rufai2020world, effenberger2020association, aiello2020social, lampos2021tracking}, economics  \cite{ettredge2005using, goel2010predicting, mclaren2011using, bollen2011twitter, choi2012predicting, preis2013quantifying, scott2014predicting,  einav2014economics,  wu2015future, vicente2015forecasting, scott2015bayesian, yi2021forecasting},  business \cite{manyika2011big, mcafee2012big, chen2012business}, finance \cite{preis2013quantifying,risteski2014can, zhu2019big}, social policy \cite{kim2014big}, popular culture trends \cite{bennett2007netflix}, among others. In particular, digital disease detection, which utilizes big data from online source to provide  accurate and up-to-date tracking of infectious diseases, has grown in popularity, especially since the 2009 H1N1 influenza pandemic and the 2020 global pandemic of COVID-19 \cite{santillana2014can, wojcik2014public, bates2017tracking, li2020retrospective, ma2022using, ma2022covid}. However, despite the existing efforts in enhancing infectious disease tracking with Internet search data, many challenges and limitations still remain. In particular, there lacks a statistically rooted and integrative approach in existing digital disease tracking frameworks that effectively accounts for the connectivity within the Internet search data. The purpose of our paper is to pioneer a statistical learning method, ARGO-C (Augmented Regression with Clustered GOogle data). ARGO-C takes advantage of the interconnections among the Internet search data and aims to improve the accuracy and interpretability of the disease tracking framework. Our method focuses on influenza (flu) tracking, but has the generality and flexibility to be adapted to tracking other infectious diseases or other social/economic trends.

Influenza (flu) epidemics occur every year with varying timing and intensity. It may claim up to 650,000 deaths per year worldwide \cite{iuliano2018estimates}, and, on average, results in 610,660 life-years loss in the US \cite{molinari2007annual}.  Our ability to prepare for and respond to epidemics or pandemics depends on the timely tracking and forecasting of the infectious disease activities \cite{lipsitch2011improving, nsoesie2014systematic, chretien2014influenza}. Traditionally, the tracking and surveillance of flu activities in the US rely mainly on the US Outpatient Influenza-like Illness Surveillance Network (ILINet) by the Centers for Disease Control (CDC). Each week, ILINet collects outpatients information from thousands of healthcare providers across the nation and reports the percentage of Influenza-like Illness patients (\%ILI). However, due to the time incurred by data collection, aggregation, and administrative processing, the CDC's weekly flu report usually lags behind real time by 1-2 weeks, which is far from optimal for tracking a fast-spreading, ever-changing disease epidemic such as flu.

In order to eliminate the time lag between CDC's flu report and the real time event, digital disease detection \cite{brownstein2009digital}, a new disease tracking framework based on Internet data, was proposed and has since revolutionized the landscape of flu tracking. In particular, methods for digital flu detection employ statistical or mechanistic models to harness Internet-derived data from various sources \cite{polgreen2008using, ginsberg2009detecting, dalton2009flutracking, achrekar2011predicting, yuan2013monitoring, lee2013real, paul2014twitter, mciver2014wikipedia, santillana2014using, paolotti2014web, smolinski2015flu, smolinski2015flu, santillana2015combining, yang2017using, bradshaw2019influenza, hassan2019social, viboud2020fitbit} to provide current estimation of future prediction of flu activity (usually in terms of \%ILI). This is also referred to as ``nowcasting", in contrast to forecasting (i.e., predicting future). Among these approaches, Google Flu Trends (GFT) \cite{ginsberg2009detecting}, which uses the volume of selected Google search terms to estimate current influenza-like illnesses (ILI) activity, has attracted the most attention. However, the significant prediction errors by GFT in the following flu season, as well as its lack of transparency and reproducibility, has incurred many criticisms. This has also inspired a growing literature \cite{cook2011assessing,  pervaiz2012flubreaks, butler2013google, olson2013reassessing, lazer2014parable, santillana2014can} in digital disease detection, with the aim to identify what Google had done wrong and improve from there.

In particular, the ARGO framework (AutoRegression with GOogle search data)  \cite{yang2015accurate} provides robust and highly accurate ILI estimates at the national level by directly addressing the limitations of GFT. Through a linear model design that is justified by a hidden Markov model, the ARGO framework effectively integrates multi-source information from the CDC's flu reports and Google's search volume data while accounting for dynamics in flu epidemics and people's search patterns.  Due to its flexibility and generality, ARGO has been well-adapted to multi-resolution, multi-disease tracking based on multi-source data \cite{yang2017using, yang2017advances, ning2019accurate, lu2019improved, yang2021use, ma2022covid, wang2022covid, ma2023joint}. 

Nevertheless, among the existing methods for digital disease tracking, few have directly addressed the complex connectivity observed within the Internet data. Particularly, many of the Internet search terms included in ARGO share semantic similarities, such as phrases like ``treat flu", ``how to treat the flu", and ``treat the flu", and consequently, their search volumes may be closely related. Such connection among Internet data has never been explicitly investigated in the existing ARGO-derived frameworks. Our goal here is to propose ARGO-C, a general framework that incorporates underlying interconnections among the Internet data and improves flu tracking's accuracy and interpretability. Our contribution is significant in that (i) ARGO-C provides an innovative statistical learning framework that explicitly models the connectivity among Internet search data and utilizes the information to improve the accuracy of disease tracking; (ii) it enhances the interpretability of the predictive model by revealing the clustering structure of search terms and each cluster's contribution to the model; (iii) it provides a general framework that is readily adaptable to tracking other diseases and/or social/economic trends with Internet search data.

\section*{Methods}
\subsection*{The ARGO model}
The model of ARGO targets the time series of logit-transformed ILI activity level $Y_t $ (i.e., the logit-transformed \%ILI) at the national-level from CDC's flu report at week $t$. It assumes a Markovian structure in a period $M$ of flu activities $\mathbf{Y}_{(t-M+1):t}$ (note that $\{(t-M+1):t\} = \{t-M+1, t-M+2, \dots, t-1, t \}$, the set of all integer indices in between), which takes the form of an autoregressive model. Consequently, $\mathbf{X}_t=(X_{1t}, X_{2t}, \dots, X_{pt})^\intercal$,  a vector of $p$ log-transformed search volumes of flu-related queries from Google, depends solely on the the flu activities at time $t$. Jointly, the structure of the search volumes $\mathbf{X}_t$ and recent flu activities $\mathbf{Y}_{(t-M+1):t}$ can be summarized by a Hidden Markov model, as presented in Equation (\ref{eq:hmm}).

\begin{equation}
\begin{array}{ccccccccc}
\cdots & \rightarrow & \mathbf{Y}_{(t-M):(t-1)} & \rightarrow & \mathbf{Y}_{(t-M+1):t}  & \rightarrow &  \mathbf{Y}_{(t-M+2):(t+1)} & \rightarrow & \cdots\\
 &  & \downarrow &  & \downarrow &  & \downarrow\\
 &  & \mathbf{X}_{t-1} &  &\mathbf{X}_{t}  &  & \mathbf{X}_{t+1} \\
\end{array}
\label{eq:hmm}
\end{equation}

With further assumptions on stationarity, normality of the observations, and linear dependence of search volumes $\mathbf{X}_{t}$ on flu activities $\mathbf{Y}_{(t-M+1):t}$, the prediction distribution $Y_t|\mathbf{Y}_{(t-M+1):(t-1)}, \mathbf{X}_{t}$ is Normal with a mean linear in $\mathbf{Y}_{(t-M+1):(t-1)}$ and $\mathbf{X}_{t}$ and a stationary variance. This leads to a linear predictive model for ARGO:

\begin{equation}
{Y}_t = \mu_y + \sum_{s=1}^M\alpha_{s}Y_{t-s} + \sum_{j=1}^p\beta_jX_{jt} + \epsilon_t,
\label{eq:argo_pred}
\end{equation}
where the random errors $\epsilon_t$ are i.i.d., with mean $0$ and constant variance $\sigma_\epsilon^2$.  Given the large number of predictors in  model \eqref{eq:argo_pred}, ARGO adopts an $L_1$-regularized regression \cite{Tibshirani96} to achieve an adaptive selection of predictors. To further account for the dynamic changes in search patterns and flu epidemics, ARGO employs a rolling window prediction scheme, with a sliding training set of $N=104$ weeks. Therefore, at a given week $T$, the coefficients $\mu_y$, $\mathbf{\beta} = (\beta_1, \beta_2, \dots, \beta_p)^\intercal$, and $\mathbf{\alpha} = (\alpha_1, \alpha_2, \dots, \alpha_M)^\intercal$ are estimated as follows:

\begin{equation}
(\hat{\mu}_y, \hat{\mathbf{\beta}}, \hat{\mathbf{\alpha}} )
    = \arg \min_{{\mu}_y, {\mathbf{\beta}}, {\mathbf{\alpha}}} 
    \left\lbrace\frac{1}{2} \sum_{t=T-N}^{T-1} (Y_t - \mu_y - \mathbf{X}_t^\intercal \mathbf{\beta})^2 
    + \lambda \| \mathbf{\beta} \|_1 
    +\lambda \| \mathbf{\alpha} \|_1 \right\rbrace,
\end{equation}
where $\|\cdot\|$ represents the $L_1$ norm, and $\lambda\ (\lambda\geq 0)$ is the tuning parameter for penalization.


\subsection*{The ARGO2 and ARGOX models for localized flu tracking}
ARGO2 \cite{ning2019accurate} generalizes the national ARGO model to localized, regional flu tracking (the US Health and Human Services (HHS) regions). It is operated in two steps:
\begin{itemize}
    \item Step One is to extract Internet search information. It employs the framework of ARGO model and applies to each region individually (with autoregressive terms left out) to obtain a preliminary raw estimate for each region's \%ILI of the week. 
    \item Step Two integrates multi-source, multi-resolution information to boost the regional \%ILI prediction. Specifically, the best linear predictor based on a structured covariance matrix is used to provide joint \%ILI prediction for all 10 regions, which incorporates Google search information from Step one's raw estimates, the national flu baseline level estimated by the ARGO model, and the recent flu time series trends from the latest CDC's flu reports. 
\end{itemize}

ARGOX \cite{yang2021use} further extends the ARGO framework to state-level, thereby establishing a coherent multi-resolution framework for digital flu tracking. Specifically, ARGOX first dichotomizes all the states into two groups: the epidemically connected and the disconnected, and then customizes different prediction models in the Step Two algorithm accordingly.

\subsection*{Penalization methods with group-wise sparsity}

As discussed in the previous sections, due to the high dimension of the Google Search terms, penalization technique is integrated into the ARGO, ARGO2, and ARGOX algorithms. In this paper, we are interested in adapting penalization methods with a group-wise regularization to these frameworks.

In linear and generalized linear regressions, penalization methods are among popular practical tools that aim at reducing the variability of parameter estimation by conventional methods, such as the ordinary least squares (OLS) method or the maximum likelihood (ML) algorithm, or decreasing the dimensionality of a given feature space. Over the past several decades, a number of penalization techniques have been developed. For instance, Horel and and Kennard\cite{Hoerl70} proposed the ridge estimator that shrinks the OLS estimator towards zero so as to alleviate its large sampling variation. Later, Lasso  and elastic net algorithms were proposed \cite{Tibshirani96,Tibshirani97,Lokhorst99, Roth04, Zou05}, both of which can realize simultaneous variable selection and parameter estimation. These methods each is defined on a different penalty function. However, they all impose regularization on individual model parameters and intend to drive each estimated parameter towards zero. 

Motivated by multi-factor analysis of variance (ANOVA), Yuan and Lin \cite{Yuan06} proposed a group lasso method that is designed to select groups of indicator variables associated with the same factor.  They combine the dummy indicator variables defined for a given factor together, and then select a subset of important factors through a penalty term imposed on the corresponding grouped model parameters. More specifically, suppose there are $p$ predictor variables which can be partitioned into $K$ groups with group size $p_k\ (1\leq k\leq K; \sum_{k=1}^K p_k=p$). In the context of linear regression, the group lasso solution for the regression coefficients $\mathbf{\beta}$ can be expressed as
\[
\hat{\mathbf{\beta}}^{\text{GL}} = \arg \min_{\mathbf{\beta}} \left\lbrace\frac{1}{2}\sum_{i=1}^n\left(Y_i-\sum_{k=1}^K (\mathbf{X}_i^{(k)})^\intercal \mathbf{\beta}^{(k)}\right)^2 + \lambda \sum_{k=1}^K \sqrt{p_k} \| \mathbf{\beta}^{(k)} \|_2 \right\rbrace,
\]
where $Y_i$ is the response variable for the $i$-th observation, $\mathbf{X}_i^{(k)}$ is the $i$-th observation's corresponding predictors in group $k$, $\mathbf{\beta}^{(k)}$ is the vector of coefficients in group $k$, and $\lambda\ (\lambda\geq 0)$ is the tuning parameter for penalization. Later, Meier et al.\cite{Meier08}
applied the group lasso method to logistic regression and other generalized linear regression models. 

More recently, Simon et al.\cite{Simon13} studied a sparse group lasso (SGL) algorithm, imposing a convex combination of the $L_1$- and $L_2$-norm penalties on the grouped and individual parameters respectively. In linear regression, the SGL solution for the regression coefficients $\mathbf{\beta}$ is
\[
\hat{\mathbf{\beta}}^{\text{SGL}} = \arg \min_{\mathbf{\beta}} \left\lbrace \frac{1}{2} \sum_{i=1}^n\left(Y_i-\sum_{k=1}^K \mathbf{X}_i^{(k)} \mathbf{\beta}^{(k)}\right)^2  + (1 - \alpha)\lambda \sum_{i=1}^m \sqrt{p_k} \|\mathbf{\beta}^{(k)}\|_2 + \alpha \lambda \|\mathbf{\beta}\|_1\right\rbrace,
\]
where $\alpha \in [0,1]$ is a weight tuning parameter, $\lambda\geq 0$, and $\|\cdot\|_1$ is the $L_1$ norm. When $\alpha=0$, the solution is reduced to the group lasso solution; when $\alpha=1$, it becomes the lasso solution. Simon et al.\cite{Simon13} also discussed the application of SGL in other model settings, such as generalized linear regression.

\subsection*{ARGO-C: the proposed method}
In this subsection, we illustrate in detail the methodology of our proposed approach, ARGO-C. 

We first recognize that some of the flu-related Google search terms often share a common theme. For example, search terms concerning flu treatments may include phrases such as ``treat flu", ``how to treat the flu", and ``treat the flu"; or a specific sub-type of flu may contain terms like ``influenza a", ``influenza type a", etc. It is then natural to consider clustering similar search terms together, and then fit a penalized linear regression model with a group-wise penalty. This direction motivated our project. 

In particular, our approach enhances the infrastructure of ARGO. Here we focus on the ILI tracking at the national level as an illustration. Additionally, to showcase the generality of our approach, we provide an exemplary integration of our approach to the ARGO2  and ARGOX frameworks \cite{ning2019accurate} for localized (regional and state-level) flu tracking. Our methodology can be readily integrated to existing methods to track  other infectious diseases \cite{yang2017advances, ma2022covid, wang2022covid, ma2023joint}, or other social and/or economic trends \cite{yi2021forecasting}.


\subsubsection*{National level}
For national level \%ILI prediction, our proposed ARGO-C is realized in two steps: in Step 1, we identify the connectivity structure among the candidate flu-related Google search terms by unsupervised statistical learning; in Step 2, we integrate the identified cluster structure into the predictive model of weekly flu activities (in terms of \%ILI) using a penalized regression with group-wise regularization.

We start with defining some notations that we will refer to for the rest of the paper. Let $Y_t$ be the logit-transformed percentage of influenza-like illness (\%ILI) at the national level at week $t$ from CDC's weekly flu reports; $\mathbf{X}_t=(X_{1t}, X_{2t}, \ldots, X_{pt})^\intercal$ be the log-transformed Google search volumes of $p$ flu-related terms at week $t$. Due to the delay of CDC's reports, at the current week $T$, we are only able to observe $Y_{1:(T-1)}$, up to the previous week; the Google search data are instead up-to-date, with $\mathbf{X}_{1:T}$ all available.  Our proposed method can be summarized as follows:
\begin{itemize}
\item \textbf{Step 1: clustering Internet search terms.}\\
Partition the $p$ search terms into $K$ groups, denoted by $G_1, \ldots, G_K$ $(K\geq 2)$. This can be done using some standard clustering method, such as hierarchical clustering \cite{Ward63}, k-means \cite{macqueen1967classification}, or model-based clustering \cite{Banerjee11}. And, standard statistical programming languages, such as \textit{R} \cite{R22} or \textit{Python} \cite{van1995python}, have existing packages that can realize these methods easily. In practice, we recommend using hierarchical clustering based on correlation as the distance metric between the search terms' time series. The number of clusters $K$ can be determined through investigating measures such as the within-group variance,  silhouette \cite{rousseeuw1987silhouettes}, or the gap statistic \cite{tibshirani2001estimating}.

\item \textbf{Step 2: nowcasting using group-structured, penalized regression.}\\
Predict the current week's (logit-transformed) ILI\%, $Y_{T}$, by
\begin{equation}
\hat{Y}_T = \mu_{y} + \sum_{s=1}^m \gamma_{s} Y_{T-s} + \sum_{j=1}^p \beta_{j}X_{jT},  \label{eq:model}
\end{equation}
where $m$ is the length of the lagged time series terms, $\mu_y$ is the intercept, $\mathbf{\gamma} = \gamma_{m:1}$ is the autoregressive coefficients, and $\mathbf{\beta} = \beta_{1:p}$ are the exogenous coefficients of Google search terms. The coefficients $\mathbf{\beta}$ are partitioned into $K$ groups as identified in the previous step. Then, the model parameters in Equation \eqref{eq:model}, $(\mu_y, \mathbf{\beta}, \mathbf{\gamma})$, can be estimated by minimizing the following penalized sum of squares quantity,

\begin{align*}
\frac{1}{2}\sum_{t=T-N}^{T-1} & 
        \Big( Y_{t} - \mu_y - \sum_{s=1}^{m} \gamma_{s} Y_{t-s}  -\sum_{j=1}^{p} \beta_{j}X_{jt}\Big) ^2
        + \alpha \lambda(\|\mathbf{\gamma}\|_1 + \|\mathbf{\beta}\|_1)  + (1 - \alpha) \lambda (\sum_{k=1}^K \sqrt{p_k} \|\mathbf{\beta}^{(k)}\|_2) \label{eq:SGL}
\end{align*}

where $N$ is the training window length, $\mathbf{\beta}^{(k)} = \{\beta_j, j\in G_k\}$ is the coefficients of search terms in cluster $k$, and $p_k$ is the size of cluster $k$. Specifically, the cluster structure of search terms is incorporated through the sparse group lasso (SGL) regularization $\|\mathbf{\beta}^{(k)}\|_2 = \sqrt{\sum_{j\in G_k} \beta_{j}^2},$
 which intends to impose sparsity on all the coefficients of terms in each cluster simultaneously. Note that $\alpha\text{ and } \lambda$ are the tuning parameters, with $\alpha$ determining the weights between the individual and group-wise regularization and $\lambda$ controlling the strength of regularization. In practice, we use SGL's default setting $\alpha=0.95$ and use cross-validation to select $\lambda$. We also follow the ARGO's default setting and set the training windows to be two years, i.e., $N=104$ weeks. 

\end{itemize}

\subsubsection*{Regional level}
For ARGO-C's regional flu tracking, we break down ARGO2's Internet data extraction step (i.e., Step 1) into two sub-steps. 

\begin{itemize}
\item \textbf{Step 1.1: clustering Internet search terms.}\\ 

For each region $r$ ($r=1,\dots, 10$), follow the similar procedure in Step 1 of the national \%ILI to partition the $p$ search terms into $K$ groups, denoted by $G_1^{(r)}, \ldots, G_K^{(r)}$. We here keep the cluster number $K$ the same as the national level for consistency.

\item \textbf{Step 1.2: extracting regional Google information based on clustered search terms.}\\
Obtain preliminary estimates for regional level \%ILI based completely on the region-wise Google search data and the clustering structure learned in Step 1.1.  Specifically, the raw estimate for region $r$'s (log-transformed) \%ILI at the current week $T$, $\hat{Y}_T^{(r)}$, is given by
\begin{equation}
\hat{Y}_T^{(r)} = \mu_{Y}^{(r)} + \sum_{j=1}^p \beta_{j}^{(r)}X_{jT}^{(r)} , \label{eq:model_reg}
\end{equation}
where the superscript $(r)$ indicates the region-specific parameters and data. And, similar to Step 2 of ARGO-C at the national level, the parameters are estimated through a sparse group lasso regularization:

\begin{align*}
\frac{1}{2N}\sum_{t=T-N}^{T-1} & 
        \Big( Y_{t}^{(r)} - \mu_y^{(r)} - \sum_{j=1}^{p} \beta_{j}^{(r)}X_{jt}^{(r)}\Big) ^2
        + \alpha \lambda\|\mathbf{\beta}^{(r)}\|_1  + (1 - \alpha) \lambda (\sum_{k=1}^K \sqrt{p_k} \|\mathbf{\beta}^{(k,r)}\|_2) \label{eq:SGL}
\end{align*}

where $N$ is the training window length, $\mathbf{\beta}^{(k,r)} = \{\beta_j^{(r)}, j\in G_k\}$ is the region-specific coefficients of search terms in cluster $k$, and $p_k$ is the size of cluster $k$.

 \item \textbf{Step 2: cross-regional boosting.} \\ This follows exactly the same as the original Step 2 in ARGO2, to prediction 10 regions' \%ILI jointly based on multi-source, multi-resolution information. More details can be found in \cite{ning2019accurate}.
\end{itemize}

\subsubsection*{State level}
At the state level, ARGO-C is readily adaptable to fit into the ARGOX framework. Step 1.1 and 1.2 follow the regional level procedure above but applied to each state; Step 2 directly inherits the original Step 2 of ARGOX \cite{yang2021use}, with the same dichotomic treatment of the 51 states/district/city.

\subsection*{Data}
\subsubsection*{CDC's \%ILI Data}
The CDC's weekly flu report is released every Friday, listing the percent of outpatient visits with influenza-like illness (\%ILI) in the \textit{previous} week \cite{CDC_FluView} (\url{https://www.cdc.gov/flu/weekly/overview.htm}). Therefore, the CDC's \%ILI always lags behind real time by at least one week. The CDC's report includes \%ILI at the national level,  of the 10 the US Health and Human Services (HHS) regions, and the 51 states/district/city (50 states plus Washington DC, excluding Florida but including New York City). The CDC’s \%ILI data for this study were collected on January 29, 2023. 

\subsubsection*{Google Search Data}
The Internet search data from Google are publicly available through Google Trends (\url{trends.google.com}). Once a user specifies a desired query term (or a topic), a geographical indicator, and a time range on Google Trends, the website will return a time series of the term's weekly search volumes. With Google Trends API, we are able to obtain the un-normalized search frequencies for the specified term, which includes all the searches that contain the entire term.


The search query terms in this study are selected based on previous works \cite{yang2015accurate, ning2019accurate, yang2021use}. Notably, we included 161 flu-related search terms/topics, with 71 terms identified by March 29, 2009 and the remaining by May 22, 2010 to account for the 2009 H1N1 outbreak. Table \ref{tab:query} lists these search terms. The regional level Google search volumes is aggregated based on state-level data, following the ARGO2 framework \cite{ning2019accurate}. 

We admit that the Google search data may only be representative of the search interests among Google users rather than the entire population. The ARGO (including ARGO2 and ARGOX) framework \cite{yang2021use} attempts to correct for such potential bias in the modeling.

As one benchmark method for comparison, we downloaded the discontinued Google Flu Trends (GFT) data (\url{https://www.google.org/flutrends/about/data/flu/us/data.txt}). GFT has the weekly \%ILI prediction from January 1, 2004 to August 9, 2015. 

\subsection*{Evaluation Metrics}

We use three metrics to evaluate the accuracy of an estimate against the actual \%ILI released by the CDC: the root mean squared error (RMSE), the mean absolute error (MAE), and the Pearson correlation (Correlation). RMSE between an estimate $\hat{p}_t$ and the true value $p_t$ over period $t=1,\ldots, N$ is given by 
\[
\text{RMSE}=\sqrt{\frac{1}{N}\sum_{t=1}^N \left(\hat{p}_t - p_t\right)^2}.
\]
The MAE between an estimate $\hat{p}_t$ and the true value $p_t$ over period $t=1,\ldots, N$ is defined as
\[
\text{MAE} = \frac{1}{N}\sum_{t=1}^N \left|\hat{p}_t - p_t\right|.
\] 
And, the correlation we considered is the Pearson correlation coefficient between $\hat{\mathbf{p}}=(\hat{p}_1, \dots, \hat{p}_N)$ and $\mathbf{p}=(p_1,\dots, p_N)$.

\section*{Results}
We first applied the ARGO-C model to retrospectively estimate the weekly \%ILI at the US national level from March 29, 2009 to January 28, 2023. Figures \ref{fig:pre09} - \ref{fig:post10} illustrate the clustering structures identified among the flu-related Google search terms by ARGO-C's Step 1. The results were obtained via hierarchical clustering with an average linkage function using correlation as the distance metric, based on the Google data available prior to the earliest prediction date. In particular, we realized the clustering analysis twice, using two sets of search terms of 71 and 161 terms/topics, respectively (see Table \ref{tab:query}). Then, the identified clusters of search terms were incorporated into ARGO-C's Step 2 to predict \%ILI after the collection date of the corresponding set of search terms (March 29, 2009 - May 21, 2010 for the first 71 terms, and May 22, 2010 onward for all 161 terms, respectively). More specifically, 53 clusters were identified among the 71 search terms originally collected by March 29, 2009 (see Figure \ref{fig:pre09}, where the number of clusters was determined by minimizing the sum of within-group variation while preserving interpretability). As can be seen, the clustering method successfully grouped search terms with close semantics into the same group. For example, one cluster consists of search terms related to flu treatments, containing phrases such as ``treat flu", ``how to treat the flu", and ``treat the flu". In addition, another cluster includes search terms about respiratory illness related to flu, such as ``sinus", ``bronchitis", ``pneumonia", and ``walking pneumonia". In the same fashion, based on the second set of 161 flu-related search terms/topics, 45 clusters were identified, as illustrated in Figure \ref{fig:post10}. 

\begin{figure}[!ht]
\centering
\includegraphics[width=\textwidth]{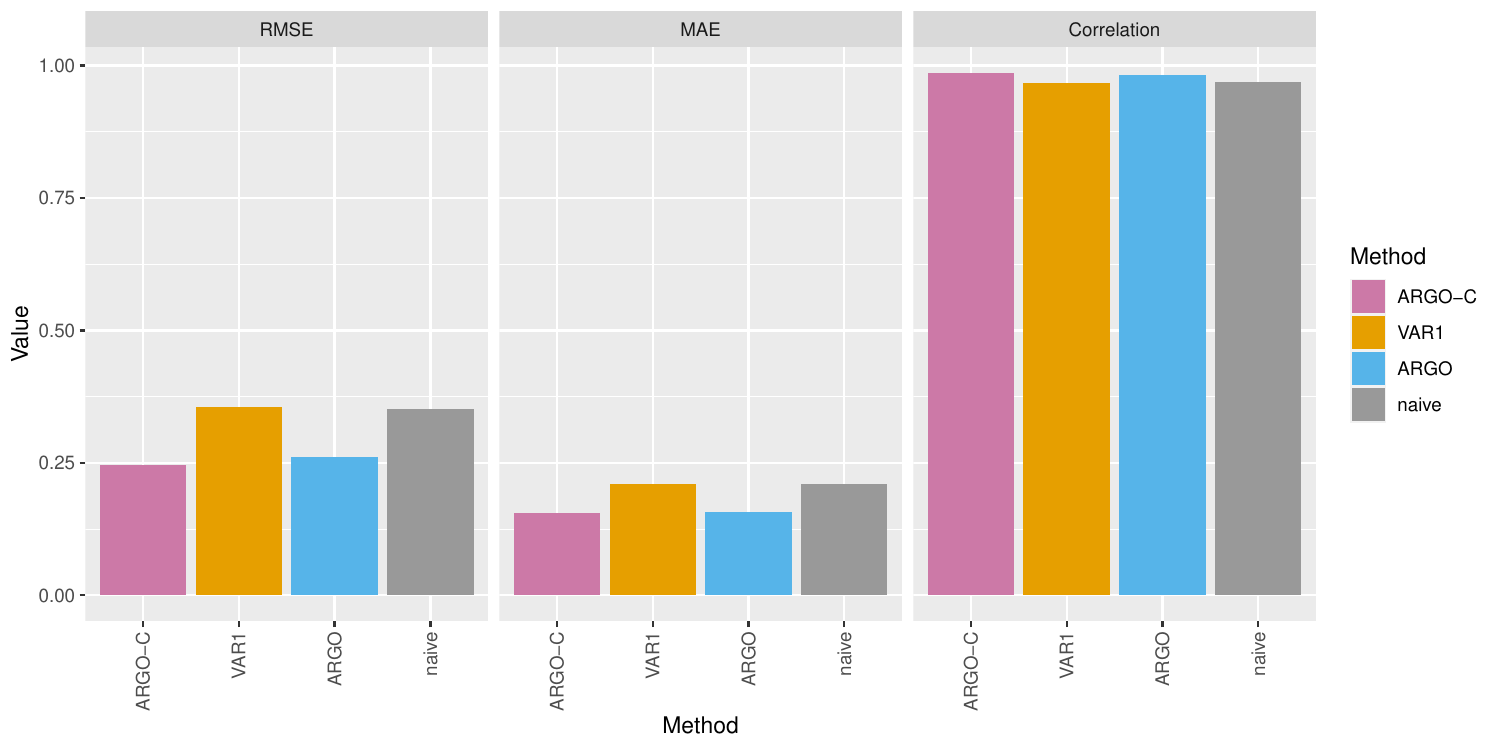}
\caption{Comparison of \%ILI estimation between ARGO-C and other benchmarks. The evaluation is based on the national level \%ILI in three accuracy metrics: RMSE, MAE, and correlation; the evaluation period is the overall period from March 29, 2009 to March 21, 2020 (excluding the influence of COVID-19). Detailed numbers can be found in Table \ref{tab:overall}. }
\label{fig:error}
\end{figure}

Figure \ref{fig:error} and Table \ref{tab:overall} summarize the national-level prediction performance of our proposed model, ARGO-C, in comparison with benchmark methods, including Google's original GFT (discontinued on July 11, 2015), vector autoregression model with lag 1 (VAR1), the original ARGO model \cite{yang2015accurate}, and the naive method which simply carries over the previous week's \%ILI to predict the current week. We first focus on the period prior to the influence of COVID-19. That is, we exclude the period when the \%ILI reported by CDC was highly confounded and contaminated by COVID-19 symptoms and cases, thus not accurately reflecting the flu activities anymore \cite{CDC_FluView, ma2023joint}. During this whole period from 2009 to 2020 (Figure \ref{fig:error}), our method ARGO-C shows the leading prediction accuracy compared to all other benchmarks across all three performance metrics. In particular, the improvement of ARGO-C from ARGO confirms the potential of integrating the interconnection among Internet search data into the modeling process while showcasing the effectiveness of ARGO-C in utilizing such information to enhance disease prediction. The effective use of connectivity among Internet data is further confirmed by a closer look at the evolving patterns of search terms included and excluded in the ARGO-C model over time. Among three exemplary clusters highlighted in Figure \ref{fig:traceplot_coef}, we observe that ARGO-C frequently selects/filters out an entire cluster of search terms thanks to the introduced group-penalty structure, and thus fully takes advantage of the interconnection among Internet data; on the other hand, each search term can also be selected/filtered out individually within a cluster, indicating a good balance between the individual and group-wise penalization. Breaking down the results into each flu season, ARGO-C's performance is consistent, giving the most accurate predictions in majority of the flu seasons. Notably, ARGO-C is the single leading method in every flu season since 2010 by the measure of correlation (Table \ref{tab:overall}). This highlights ARGO-C's strength in predicting the flu epidemic trends. Moreover, ARGO-C's advantage over the vanilla ARGO is significantly evident in certain difficult flu seasons, yielding up to 30\% of error reduction (i.e., there are error reductions of 30.7\% for '10-'11 ,  18.4\% for '14-'15, and 18.9\%  for '16-'17 in terms of RSME). Additionally, the 95\% prediction interval given by ARGO-C (based on the stationary bootstrap \cite{yang2015accurate}) has an empirical coverage of 95.08\%.   

We also applied the proposed ARGO-C model to localized flu tracking. At the regional level,  ARGO-C again shows the strongest performance across all 10 regions in all three accuracy metrics. More specifically, compared to the naive method, ARGO-C reduces the RMSE by 12\% to 29\%, and improves the MAE measure by 11\% to 25\% (Table \ref{tab:region}). In addition, the correlation measure based on the ARGO-C method is uniformly higher than other benchmarks in all 10 regions. Breaking down into individual flu seasons (Table \ref{tab:reg1}-\ref{tab:reg10}), ARGO-C still leads in  vast majority of all evaluated periods. The strengths of ARGO-C at the regional level also reaffirms our projection that interconnection among Internet search data would contribute effectively in improving flu tracking performance. Additionally, Table \ref{tab:ci_region} reports the empirical interval coverage for the regional \%ILI prediction (ranging from 93\% to 96\%, given a nominal level of 95\%), further confirming the reliability of ARGO-C.

Table \ref{tab:state} and \ref{tab:state_covid} summarize ARGO-C's state-level flu tracking performance in comparison with the benchmarks. Averaging over 51 states, ARGO-C again is the best performing method compared to all benchmarks, showcasing its adaptability and robustness in high-resolution disease tracking. The strength of ARGO-C attributes to our effective modeling of interconnection among search terms, which is even more relevant for efficient extracting of low-quality, high-noise Internet data at high resolution. More detailed reports on each individual state and each flu season are given in Tables \ref{tab:state US.AL} - \ref{tab:state US.NYC}. During the period from 2014 to 2020 (pre-COVID, Table \ref{tab:state}), ARGO-C leads the chart for the majority of the states (ARGO-C outperforms other models for 42 states in terms of MSE). Notably, after including the irregular flu seasons since COVID (2014-2023, Table \ref{tab:state_covid}), ARGO-C shows even more remarkable strength over benchmarks (ARGO-C outperforms other models for 47 states in terms of MSE). To further confirm the reliability of ARGO-C, we report the actual coverage rate of the 95\% prediction interval given by ARGO-C for each state in Table \ref{tab:ci_state}: Overall, the average coverage rate over the 51 states is 92.6\%, close to the nominal level.

\section*{Discussion}
In summary, to account for the interconnection among the Internet search data, we proposed an innovative statistical learning framework, ARGO-C. By applying ARGO-C to both national and localized flu tracking, we observe that ARGO-C enhances the original ARGO/ARGO2/ARGOX framework by effectively and efficiently extracting and utilizing the inherent grouping structure of Google search terms.

The first step model of our proposed ARGO-C identifies the interconnection structure among Google search terms through clustering. In general, any clustering technique may be employed at this stage, and for each given method, several fitting criteria may also be considered, such as different choices of the distance metric and/or other tuning parameters (e.g., linkage function for hierarchical clustering). Specifically for the task of flu prediction, we recommend using hierarchical clustering based on the correlation distance metric and the average linkage function. This is through our empirical explorations of multiple classic clustering methods, including hierarchical clustering, k-means, and PAM \cite{kaufman1990partitioning} with various tuning configurations (linkage, distance metric, etc.)  In practice, one may also explore a few different clustering methods and choose the one that gives the most interpretable results in the given context. It is also possible to realize clustering in a more dynamic fashion, updating the search term clusters periodically over time, which may possibly improve the accuracy of the \%ILI predictions in the following modeling step. In addition, we also explored various criteria to determine the number of clusters $K$. Our final choice of $K$ was a joint decision based on the within-cluster distance with consideration of model interpretability. In practice, we recommend choosing a relatively large $K$ for effective incorporation of the clustering information. 

We acknowledge that there are limitations of utilizing Google search data into \%ILI predictions during the COVID-19 pandemic or the post-COVID period in the near future. First of all, \%ILI is only a proxy for the actual flu incidence in the population. Since the seasonal flu and COVID-19 share many symptoms in common, the reported \%ILI may as well include visits due to COVID-19. Consequently, the \%ILI predictions may be largely influenced by simultaneous COVID-19 seasonal surges or underlying COVID-19 cases. In the main focus of our data analysis, we excluded the period that was possibly contaminated by COVID-19, as we suspect that the \%ILI target and the existing set of Google search terms may not well represent flu activities during that period. Nevertheless, the \%ILI surveillance data can still provide valuable insights on  the general trend of influenza activity \cite{CDC_FluView} (so, we also presented the results during the post-COVID period in the Supplementary Information, which also showcases the robustness of ARGO-C). It could be among our future endeavors to update the Google search terms after accounting for the effect of COVID-19, and/or to target alternative flu indicators \cite{chretien2014influenza}, such as laboratory-confirmed influenza hospital admissions \cite{CDC_FluSight}. In addition, it will also be an interesting future project to explore the possibility of predicting seasonal flu cases and COVID-19 cases simultaneously by accounting for their interactive effects on each other \cite{ma2023joint}.

Although not presented in this paper, ARGO-C is highly robust and can be easily adapted to digital tracking of other diseases or social/economic trends (as we have done to ILI prediction at various geographical resolution). We hope that our proposed framework can improve the real-time tracking of various infectious diseases and potentially contribute to the area of public health by saving more people's lives.

\section*{Acknowledgements}
The work by A.H. was supported by the Finnerty Fund for summer undergraduate research in the Department of Mathematics and Statistics, Williams College, and by the College's DFRC fund for undergraduate research assistant.

\section*{Author contributions statement}

S.N. and Q.W. conceived the research; S.N. and A.H. conducted the data analysis and experiments; S.N., A.H., and Q.W. analyzed the results; S.N. and Q.W. wrote the paper. All authors reviewed the manuscript. 

\section*{Data availability statement}
The datasets generated and analysed during the current study, as well as the codes generating the results, are available at \url{https://github.com/shaoyangning/argo-c}.

\section*{Additional information}
\subsection*{Competing Interests}
The authors declare no competing interests.

\bibliography{mybibfile}

\begin{table}[ht]
\begin{adjustwidth}{-4.5cm}{-4cm}
\small
\centering
\begin{tabular}{c|lllllllllllll}
  \hline
 & Whole period & GFT period & '09-'10 & '10-'11 & '11-'12 & '12-'13 & '13-'14 & '14-'15 & '15-'16 & '16-'17 & '17-'18 & '18-'19 & '19-'20 \\ 
 \hline
        \multicolumn{1}{l|}{RSME} \\
ARGO-C & \textbf{0.246} & \textbf{0.283} & 0.457 & \textbf{0.233} & 0.153 & 0.489 & \textbf{0.152} & \textbf{0.206} & \textbf{0.149} & \textbf{0.270} & 0.204 & \textbf{0.171} & \textbf{0.425} \\ 
  GFT & -- & 0.770 & \textbf{0.437} & 0.376 & 0.493 & 2.221 & 0.346 & 0.316 & -- & -- & -- & -- & -- \\ 
  VAR1 & 0.355 & 0.348 & 0.521 & 0.333 & 0.164 & 0.503 & 0.353 & 0.455 & 0.237 & 0.364 & 0.561 & 0.346 & 0.695 \\ 
  ARGO & 0.261 & 0.294 & 0.454 & 0.336 & \textbf{0.132} & \textbf{0.455} & 0.184 & 0.244 & 0.159 & 0.333 & \textbf{0.179} & 0.204 & 0.436 \\ 
  naive & 0.352 & 0.347 & 0.520 & 0.339 & 0.163 & 0.499 & 0.350 & 0.457 & 0.234 & 0.343 & 0.556 & 0.344 & 0.691 \\ 
   \hline
     \hline
       \multicolumn{1}{l|}{MAE} \\
ARGO-C & \textbf{0.156} & 0.178 & 0.300 & \textbf{0.176} & 0.126 & 0.306 & \textbf{0.119} & \textbf{0.153} & \textbf{0.108} & \textbf{0.188} & 0.152 & \textbf{0.122} & \textbf{0.297} \\ 
  GFT & -- & 0.366 & 0.294 & 0.327 & 0.441 & 1.634 & 0.189 & 0.214 & -- & -- & -- & -- & -- \\ 
  VAR1 & 0.210 & 0.199 & 0.322 & 0.256 & 0.136 & 0.314 & 0.211 & 0.283 & 0.191 & 0.267 & 0.377 & 0.280 & 0.538 \\ 
  ARGO & 0.158 & \textbf{0.172} & \textbf{0.290} & 0.265 & \textbf{0.104} & \textbf{0.260} & 0.120 & 0.178 & 0.135 & 0.218 & \textbf{0.141} & 0.149 & 0.334 \\ 
  naive & 0.210 & 0.201 & 0.323 & 0.259 & 0.135 & 0.325 & 0.212 & 0.290 & 0.187 & 0.256 & 0.384 & 0.282 & 0.533 \\ 
   \hline
    \hline
           \multicolumn{1}{l|}{Correlation} \\
ARGO-C & \textbf{0.985} & \textbf{0.977} & 0.982 & \textbf{0.980} & \textbf{0.922} & \textbf{0.959} & \textbf{0.987} & \textbf{0.988} & \textbf{0.981} & \textbf{0.978} & \textbf{0.997} & \textbf{0.992} & \textbf{0.975} \\ 
  GFT & -- & 0.876 & \textbf{0.995} & 0.968 & 0.833 & 0.926 & 0.969 & 0.987 & -- & -- & -- & -- & -- \\ 
  VAR1 & 0.968 & 0.960 & 0.967 & 0.955 & 0.885 & 0.922 & 0.921 & 0.938 & 0.932 & 0.950 & 0.965 & 0.956 & 0.933 \\ 
  ARGO & 0.983 & 0.971 & 0.975 & 0.963 & \textbf{0.922} & 0.940 & 0.978 & 0.986 & 0.970 & \textbf{0.978} & \textbf{0.997} & 0.988 & 0.974 \\ 
  naive & 0.968 & 0.961 & 0.968 & 0.954 & 0.887 & 0.924 & 0.923 & 0.939 & 0.935 & 0.956 & 0.964 & 0.957 & 0.933 \\ 
   \hline
\end{tabular}
\end{adjustwidth}
\caption{Comparison of national \%ILI estimation between ARGO-C and other benchmarks. The evaluation is based on the national level \%ILI in multiple periods and multiple metrics. RMSE, MAE, and correlation are reported. The method with the best performance is highlighted in boldface for each metric in each period. Methods considered here include ARGO-C, VAR1, GFT, the original ARGO, and the naive method. All comparisons are conducted on the original scale of the CDC’s \%ILI. The whole period is March 29, 2009 to March 21, 2020, excluding the period with COVID influence. The second column (GFT period) is the period when the estimation by GFT is available, i.e., March 29, 2009 to July 11, 2015. The remaining columns contain yearly regular flu seasons, from week 40 to week 20 next year, as defined by CDC’s Morbidity and Mortality Weekly Report. (The 19’-20’ season is up to March 21, 2020).} \label{tab:overall}
\end{table}

\begin{table}[!ht]
\begin{adjustwidth}{-4.5cm}{-4cm}
\small
\centering
\begin{tabular}{c|llllllllll}
  \hline
  & Region 1 & Region 2 & Region 3 & Region 4 & Region 5 & Region 6 & Region 7 & Region 8 & Region 9 & Region 10 \\ 
  \hline \hline \multicolumn{1}{l|}{RMSE}\\ARGO-C & \textbf{0.301} & \textbf{0.399} & \textbf{0.338} & \textbf{0.334} & \textbf{0.281} & \textbf{0.584} & \textbf{0.423} & \textbf{0.310} & \textbf{0.364} & \textbf{0.372} \\ 
  ARGO2 & 0.308 & 0.410 & 0.367 & 0.347 & 0.302 & 0.604 & 0.454 & 0.319 & 0.376 & 0.382 \\ 
  VAR1 & 0.369 & 0.487 & 0.476 & 0.462 & 0.391 & 0.730 & 0.565 & 0.382 & 0.413 & 0.438 \\ 
  naive & 0.366 & 0.487 & 0.473 & 0.460 & 0.389 & 0.722 & 0.558 & 0.374 & 0.415 & 0.432 \\ 
   \hline \hline \multicolumn{1}{l|}{MAE}\\ARGO-C & \textbf{0.162} & \textbf{0.267} & \textbf{0.210} & \textbf{0.203} & \textbf{0.159} & \textbf{0.353} & \textbf{0.247} & \textbf{0.175} & \textbf{0.224} & \textbf{0.228} \\ 
  ARGO2 & 0.164 & 0.275 & 0.226 & 0.212 & 0.167 & 0.371 & 0.266 & 0.180 & 0.229 & 0.230 \\ 
  VAR1 & 0.191 & 0.303 & 0.266 & 0.264 & 0.208 & 0.440 & 0.313 & 0.206 & 0.253 & 0.247 \\ 
  naive & 0.192 & 0.304 & 0.267 & 0.266 & 0.211 & 0.440 & 0.316 & 0.205 & 0.255 & 0.256 \\ 
   \hline \hline \multicolumn{1}{l|}{Correlation}\\ARGO-C & \textbf{0.965} & \textbf{0.969} & \textbf{0.971} & \textbf{0.978} & \textbf{0.972} & \textbf{0.973} & \textbf{0.972} & \textbf{0.973} & \textbf{0.946} & \textbf{0.955} \\ 
  ARGO2 & 0.964 & 0.968 & 0.967 & 0.976 & 0.969 & 0.972 & 0.969 & 0.972 & 0.943 & 0.953 \\ 
  VAR1 & 0.946 & 0.954 & 0.941 & 0.957 & 0.945 & 0.957 & 0.948 & 0.958 & 0.931 & 0.938 \\ 
  naive & 0.948 & 0.955 & 0.943 & 0.958 & 0.946 & 0.959 & 0.950 & 0.960 & 0.932 & 0.941 \\ 
   \hline
\end{tabular}
\end{adjustwidth}
\caption{Comparison of regional \%ILI estimation between ARGO-C and other benchmarks. The evaluation is based on the \%ILI at 10 HHS regional level in multiple metrics. RMSE, MAE, and correlation are reported. The method with the best performance is highlighted in boldface for each metric in each period. Methods considered here include ARGO-C, VAR1, the original ARGO2, and the naive method. All comparisons are conducted on the original scale of the CDC’s \%ILI. The evaluation period is March 29, 2009 to March 21, 2020, excluding the period with COVID-19 influence.} \label{tab:region}
\end{table}

\begin{table}[!ht]
\begin{adjustwidth}{-4.5cm}{-4cm}
\small
\centering
\begin{tabular}{rrrrrrrrrr}
  \hline
  Region 1 & Region 2 & Region 3 & Region 4 & Region 5 & Region 6 & Region 7 & Region 8 & Region 9 & Region 10 \\ 
  \hline
  0.935 & 0.930 & 0.953 & 0.946 & 0.953 & 0.942 & 0.958 & 0.939 & 0.944 & 0.951 \\ 
   \hline
\end{tabular}
\end{adjustwidth}
\caption{Actual coverage of prediction intervals by ARGO-C for regional \%ILI prediction. The coverage is for 95\% nominal level. The average coverage over the ten US HHS regions is 94.5\%. The evaluation period is March 29, 2009 to March 21, 2020, excluding the period with COVID-19 influence.} \label{tab:ci_region}
\end{table}

\begin{table}[!ht]
\small
\centering
\begin{tabular}{c|lllllll}
  \hline
  & Whole period & GFT period & '15-'16 & '16-'17 & '17-'18 & '18-'19 & '19-'20 \\ 
  \hline \hline \multicolumn{1}{l|}{RMSE}\\ARGO-C & \textbf{0.569} & \textbf{0.714} & \textbf{0.473} & \textbf{0.662} & 0.680 & \textbf{0.548} & \textbf{0.877} \\ 
  ARGOX & 0.578 & 0.733 & 0.474 & 0.673 & \textbf{0.680} & 0.553 & 0.936 \\ 
  VAR1 & 1.855 & 2.300 & 1.350 & 1.992 & 2.703 & 1.822 & 2.785 \\ 
  GFT & -- & 1.546 & -- & -- & -- & -- & -- \\ 
  naive & 0.680 & 0.849 & 0.507 & 0.743 & 0.894 & 0.659 & 1.063 \\ 
   \hline \hline \multicolumn{1}{l|}{MAE}\\ARGO-C & \textbf{0.336} & \textbf{0.402} & 0.321 & \textbf{0.412} & 0.440 & \textbf{0.357} & \textbf{0.561} \\ 
  ARGOX & 0.339 & 0.410 & \textbf{0.318} & 0.421 & \textbf{0.438} & 0.362 & 0.590 \\ 
  VAR1 & 1.186 & 1.489 & 1.032 & 1.396 & 1.738 & 1.277 & 2.044 \\ 
  GFT & -- & 1.001 & -- & -- & -- & -- & -- \\ 
  naive & 0.388 & 0.467 & 0.340 & 0.464 & 0.558 & 0.443 & 0.691 \\ 
   \hline \hline \multicolumn{1}{l|}{Correlation}\\ARGO-C & \textbf{0.950} & \textbf{0.911} & 0.821 & \textbf{0.872} & 0.934 & \textbf{0.921} & \textbf{0.905} \\ 
  ARGOX & 0.948 & 0.907 & \textbf{0.824} & 0.869 & \textbf{0.935} & 0.920 & 0.896 \\ 
  VAR1 & 0.728 & 0.696 & 0.636 & 0.660 & 0.785 & 0.759 & 0.700 \\ 
  GFT & -- & 0.902 & -- & -- & -- & -- & -- \\ 
  naive & 0.930 & 0.877 & 0.803 & 0.842 & 0.899 & 0.890 & 0.873 \\ 
   \hline
\end{tabular}
\caption{Comparison of state-level \%ILI estimation between ARGO-C and other benchmarks. The evaluation is based on the average of 51 US state/district in multiple periods and multiple metrics. RMSE, MAE, and correlation are reported. The method with the best performance is highlighted in boldface for each metric in each period. Methods considered here include ARGO-C, VAR1, GFT, the original ARGOX, and the naive method. All comparisons are conducted on the original scale of the CDC’s \%ILI. The whole period is January 10, 2014 (first available estimate by ARGOX) to March 21, 2020, excluding the period with COVID influence. The second column (GFT period) is the period when the estimation by GFT is available, i.e., January 10, 2014 to July 11, 2015. Each regular flu season ('14-'15 overlaps with GFT period) is from week 40 to week 20 next year, as defined by CDC’s Morbidity and Mortality Weekly Report. (The 19’-20’ season is up to March 21, 2020).} \label{tab:state}
\end{table}

\label{myLastPage}
\newpage

\include{supplement}

\include{tables_state_error}

\end{document}

%% file: supplement.tex
\setcounter{page}{1}

\rfoot{\thepage}
\section*{Supplementary Information}

\setcounter{table}{0}
\renewcommand{\thetable}{S\arabic{table}}%
\setcounter{figure}{0}
\renewcommand{\thefigure}{S\arabic{figure}}%

\begin{sidewaysfigure}[htbp]
\begin{adjustwidth}{-4.5cm}{-4.5cm}
\centering
\includegraphics[width=\textwidth]{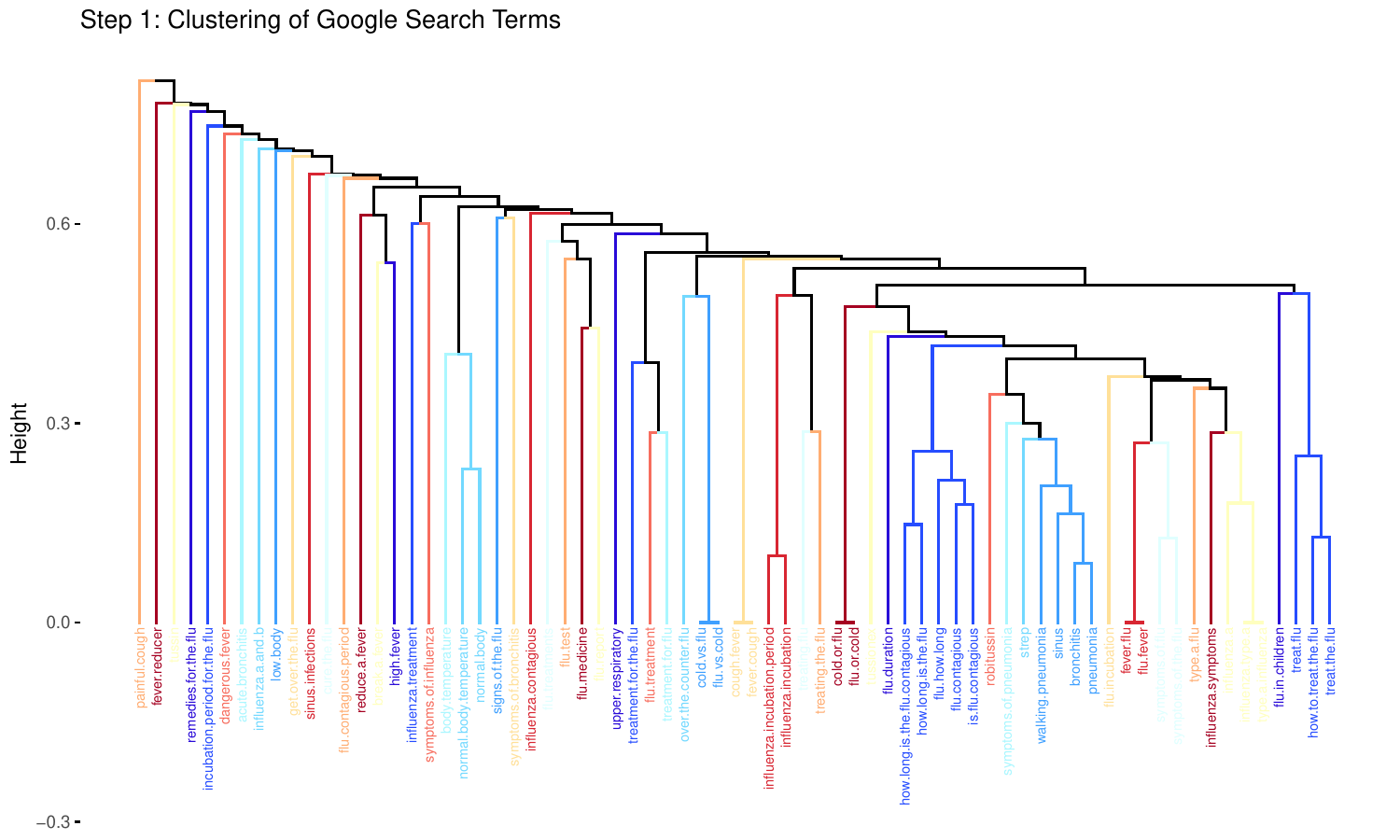}
\end{adjustwidth}
\caption{Clustering of the 71 Google search terms, collected by March 29, 2009. 53 clusters were identified. Hierarchical clustering with average linkage and correlation distance metric were used. The clustering was conducted based on the time series of Google search data from January 10, 2004 (the earliest available Google Trends data) to March 29, 2009 (the earliest prediction date by these 71 terms).}
\label{fig:pre09}
\end{sidewaysfigure}

\begin{sidewaysfigure}[htbp]
\begin{adjustwidth}{-4.5cm}{-4.5cm}
\centering
\includegraphics[width=1\textwidth]{{figs/fig\_clust\_hc\_corr\_ave\_post10}.pdf}
\end{adjustwidth}
\caption{Clustering of the 161 Google search terms, collected by May 22, 2010. 45 clusters were identified. Hierarchical clustering with average linkage and correlation distance metric were adopted. The clustering was conducted based on the time series of Google search data from January 10, 2004 (the earliest available Google Trends data) to May 22, 2010 (the earliest prediction date by these 161 terms).}
\label{fig:post10}
\end{sidewaysfigure}

\begin{figure}[!ht]
\centering
\includegraphics[width=0.6\textwidth]
{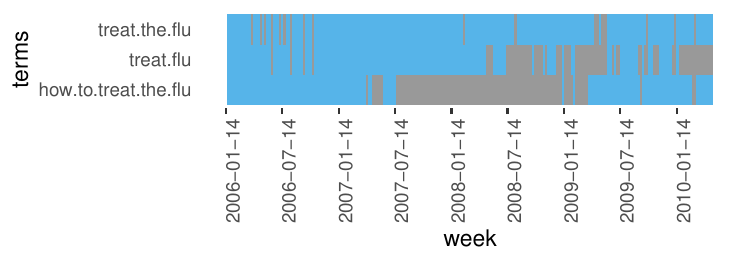}
\includegraphics[width=0.6\textwidth]
{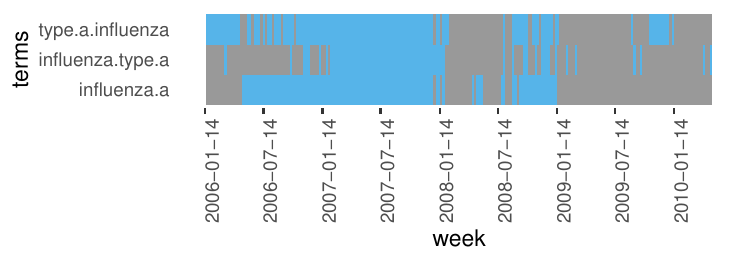}
\includegraphics[width=\textwidth]
{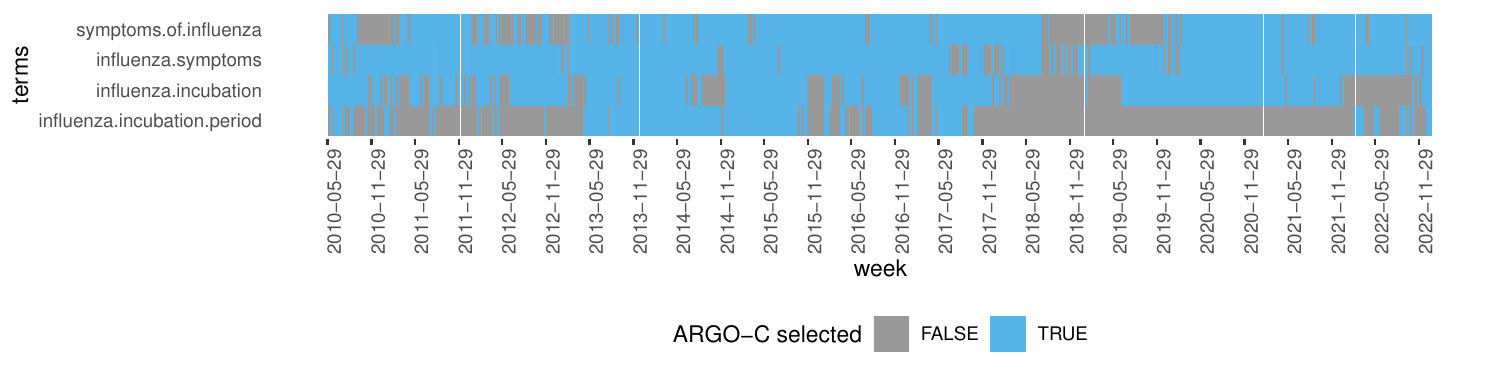}
\caption{Traceplots of clustered predictors included in the ARGO-C model at the national level over time. The heatmaps indicate whether each predictor was included in the predictive ARGO-C model at each week. Three exemplary clusters were highlighted. Each of the top two clusters contains three search terms (identified among 71 search terms by March 29, 2009, used for predictions before 2010), while the last one includes four search terms (identified among 161 topics/terms by May 22, 2010 and used for predictions since 2010). The entire cluster was penalized and excluded from the model when an entire column in the traceplot is colored grey.}\label{fig:traceplot_coef}
\end{figure}

\begin{table}[!ht]
\begin{adjustwidth}{-4.5cm}{-4cm}
\small
\centering
\begin{tabular}{c|llll}
  \hline
& Overall '09-'23 & '20-'21 & '21-'22 & '22-'23 \\ 
  \hline          
  \multicolumn{1}{l|}{RSME} \\
ARGO-C & \textbf{0.252} & 0.113 & \textbf{0.224} & 0.490 \\ 
  GFT & -- & -- & -- & -- \\ 
  VAR1 & 0.359 & \textbf{0.081} & 0.357 & 0.769 \\ 
  ARGO & 0.258 & 0.100 & 0.232 & \textbf{0.487} \\ 
  naive & 0.355 & 0.082 & 0.358 & 0.734 \\ 
   \hline
   \hline  \multicolumn{1}{l|}{MAE}\\
ARGO-C & 0.161 & 0.089 & 0.181 & 0.429 \\ 
  GFT & -- & -- & -- & -- \\ 
  VAR1 & 0.211 & \textbf{0.066} & 0.245 & 0.644 \\ 
  ARGO & \textbf{0.158} & 0.086 & \textbf{0.168} & \textbf{0.374} \\ 
  naive & 0.210 & 0.067 & 0.239 & 0.591 \\ 
   \hline  \multicolumn{1}{l|}{Correlation}\\ARGO-C & \textbf{0.984} & 0.927 & \textbf{0.983} & \textbf{0.975} \\ 
  GFT & -- & -- & -- & -- \\ 
  VAR1 & 0.968 & \textbf{0.941} & 0.899 & 0.909 \\ 
  ARGO & 0.983 & 0.940 & 0.962 & 0.969 \\ 
  naive & 0.968 & 0.939 & 0.900 & 0.907 \\ 

   \hline
\end{tabular}
\end{adjustwidth}
\caption{Comparison of \% ILI estimation between ARGO-C and other benchmarks at the national level, for flu seasons since COVID-19. The evaluation is based at the national level \%ILI in multiple periods and multiple metrics. RMSE, MAE, and correlation are reported. The method with the best performance is highlighted in boldface for each metric in each period. Methods considered here include ARGO-C, VAR1, GFT, the original ARGO, and the naive method. All comparisons are conducted on the original scale of the CDC’s \%ILI. The overall period '09-'23 is March 29, 2009 to January 28, 2023, including the period since COVID. Each regular flu season is from week 40 to week 20 next year, as defined by CDC’s Morbidity and Mortality Weekly Report. (The '22-'23 season is up to January 28, 2023).} \label{tab:covid}
\end{table}

\begin{table}[!ht]
\small
\centering
\begin{tabular}{c|lllll}
  \hline
  & Overall '14-'23 & post-COVID & '20-'21 & '21-'22 & '22-'23 \\ 
  \hline \hline \multicolumn{1}{l|}{RMSE}\\ARGO-C & \textbf{0.564} & \textbf{0.554} & 0.287 & \textbf{0.480} & \textbf{1.048} \\ 
  ARGOX & 0.576 & 0.574 & 0.283 & 0.488 & 1.101 \\ 
  VAR1 & 1.789 & 1.660 & 0.736 & 1.368 & 3.398 \\ 
  GFT & -- & -- & -- & -- & -- \\ 
  naive & 0.662 & 0.626 & \textbf{0.279} & 0.551 & 1.224 \\ 
   \hline \hline \multicolumn{1}{l|}{MAE}\\ARGO-C & \textbf{0.325} & \textbf{0.304} & 0.186 & 0.306 & \textbf{0.685} \\ 
  ARGOX & 0.329 & 0.309 & 0.182 & \textbf{0.305} & 0.735 \\ 
  VAR1 & 1.114 & 0.979 & 0.511 & 0.954 & 2.364 \\ 
  GFT & -- & -- & -- & -- & -- \\ 
  naive & 0.370 & 0.336 & \textbf{0.166} & 0.346 & 0.849 \\ 
   \hline \hline \multicolumn{1}{l|}{Correlation}\\ARGO-C & \textbf{0.947} & \textbf{0.933} & 0.703 & \textbf{0.862} & \textbf{0.890} \\ 
  ARGOX & 0.944 & 0.929 & 0.709 & 0.855 & 0.877 \\ 
  VAR1 & 0.675 & 0.719 & 0.478 & 0.563 & 0.591 \\ 
  GFT & -- & -- & -- & -- & -- \\ 
  naive & 0.929 & 0.919 & \textbf{0.710} & 0.816 & 0.844 \\ 
   \hline
\end{tabular}
\caption{Comparison of \% ILI estimation between ARGO-C and other benchmarks at the state level, for flu seasons since COVID-19. The evaluation is based on the average of 51 US state/district in multiple periods and multiple metrics. RMSE, MAE, and correlation are reported. The method with the best performance is highlighted in boldface for each metric in each period. Methods considered here include ARGO-C, VAR1, GFT, the original ARGOX, and the naive method. All comparisons are conducted on the original scale of the CDC’s \%ILI. The overall period '14-'23 is January 10, 2014 (first available estimate by ARGO framework) to January 28, 2023, including the period since COVID. The post-COVID period is the period since COVID, March 21, 2020 to  January 28, 2023. Each regular flu season is from week 40 to week 20 next year, as defined by CDC’s Morbidity and Mortality Weekly Report. (The '22-'23 season is up to January 28, 2023).} \label{tab:state_covid}
\end{table}

\begin{table}[!ht]
\centering
\begin{tabular}{rrrrrrrrrrrrrrr}
  \hline
AL & AK & AZ & AR & CA & CO & CT & DE & DC & GA & HI & ID & IL & IN & IA \\ 
0.931 & 0.926 & 0.915 & 0.929 & 0.947 & 0.903 & 0.906 & 0.947 & 0.908 & 0.933 & 0.942 & 0.933 & 0.931 & 0.924 & 0.931 \\ 
   \hline
   \hline
  KS & KY & LA & ME & MD & MA & MI & MN & MS & MO & MT & NE & NV & NH & NJ \\ 

0.938 & 0.894 & 0.922 & 0.917 & 0.933 & 0.919 & 0.940 & 0.942 & 0.938 & 0.935 & 0.935 & 0.910 & 0.903 & 0.908 & 0.929 \\    \hline
  \hline
NM & NY & NC & ND & OH & OK & OR & PA & RI & SC & SD & TN & TX & UT & VT \\ 

0.915 & 0.899 & 0.952 & 0.949 & 0.924 & 0.924 & 0.926 & 0.940 & 0.919 & 0.938 & 0.940 & 0.915 & 0.933 & 0.931 & 0.945 \\ 
   \hline
     \hline
VA & WA & WV & WI & WY & NYC \\ 
0.954 & 0.908 & 0.917 & 0.945 & 0.885 & 0.915 \\ 
   \hline
\end{tabular}
\caption{Actual coverage of prediction intervals by ARGO-C for state-level \%ILI prediction. The coverage is for 95\% nominal level. The average coverage over 51 states/city/district is 92.6\%. The evaluation period is January 10, 2014 to March 21, 2020, excluding the period with COVID-19 influence.} \label{tab:ci_state}
\end{table}

\begin{table}[!ht]
\begin{adjustwidth}{0cm}{0cm}
\footnotesize
\centering
\begin{tabular}{llll}
  \hline
    acute bronchitis & body temperature & break a fever & bronchitis \\ 
  cold or flu & cold vs flu & cough fever & cure the flu \\ 
  dangerous fever & fever cough & fever flu & fever reducer \\ 
  flu contagious period & flu contagious & flu duration & flu fever \\ 
  flu how long & flu in children & flu incubation & flu medicine \\ 
  flu or cold & flu report & flu test & flu treatment \\ 
  flu treatments & flu vs cold & get over the flu & high fever \\ 
  how long is the flu contagious & how long is the flu & how to treat the flu & incubation period for the flu \\
\hline
\end{tabular}
\caption{All search query terms used in this study. The first 71 terms were collected on March 29, 2009, the remaining terms/topics were identified on May 22, 2010. The last 21 terms separated by a horizontal line from the first 140 terms were ``Related topics'' from Google Trends.}\label{tab:query}
\end{adjustwidth}
\end{table}

\begin{table*}
\begin{adjustwidth}{0cm}{0cm}
\tiny
\centering
\begin{tabular}{c|llllllllllllllll}
  \hline
  & '09-'23 & '09-'15 & '09-'10 & '10-'11 & '11-'12 & '12-'13 & '13-'14 & '14-'15 & '15-'16 & '16-'17 & '17-'18 & '18-'19 & '19-'20 & '20-'21 & '21-'22 & '22-'23 \\ 
  \hline \hline \multicolumn{1}{l}{RMSE}\\ARGO-C & \textbf{0.307} & \textbf{0.349} & 0.903 & 0.174 & 0.115 & \textbf{0.253} & \textbf{0.214} & \textbf{0.321} & \textbf{0.183} & \textbf{0.214} & \textbf{0.256} & \textbf{0.180} & \textbf{0.432} & 0.092 & \textbf{0.297} & \textbf{0.784} \\ 
  ARGO2 & 0.316 & 0.353 & 0.904 & \textbf{0.172} & 0.104 & 0.284 & 0.216 & 0.330 & 0.190 & 0.225 & 0.273 & 0.187 & 0.450 & 0.096 & 0.306 & 0.829 \\ 
  VAR1 & 0.382 & 0.409 & 1.069 & 0.194 & \textbf{0.104} & 0.372 & 0.214 & 0.371 & 0.212 & 0.276 & 0.440 & 0.271 & 0.636 & \textbf{0.067} & 0.365 & 1.011 \\ 
  GFT & -- & 0.941 & \textbf{0.466} & 0.237 & 0.245 & 2.819 & 0.395 & 0.418 & -- & -- & -- & -- & -- & -- & -- & -- \\ 
  naive & 0.377 & 0.406 & 1.062 & 0.193 & 0.105 & 0.362 & 0.214 & 0.360 & 0.213 & 0.264 & 0.436 & 0.271 & 0.628 & 0.068 & 0.360 & 0.970 \\ 
   \hline \hline \multicolumn{1}{l}{MAE}\\
\hline
\end{tabular}
\caption{Comparison of different methods for regional level \%ILI estimation in Region 1. The RMSE, MAE, and correlation measures are reported. The method with the best performance is highlighted in boldface for each metric in each period. \label{tab:reg1}} 
\end{adjustwidth}\end{table*}

\begin{table*}
\begin{adjustwidth}{0cm}{0cm}
\tiny
\centering
\begin{tabular}{c|llllllllllllllll}
  \hline
  & '09-'23 & '09-'15 & '09-'10 & '10-'11 & '11-'12 & '12-'13 & '13-'14 & '14-'15 & '15-'16 & '16-'17 & '17-'18 & '18-'19 & '19-'20 & '20-'21 & '21-'22 & '22-'23 \\ 
  \hline \hline \multicolumn{1}{l}{RMSE}\\ARGO-C & \textbf{0.446} & \textbf{0.418} & 0.646 & \textbf{0.385} & 0.134 & \textbf{0.340} & 0.337 & \textbf{0.374} & 0.354 & \textbf{0.540} & 0.599 & \textbf{0.228} & 0.953 & 0.229 & \textbf{0.552} & \textbf{0.885} \\ 
  ARGO2 & 0.458 & 0.430 & 0.647 & 0.394 & 0.125 & 0.395 & \textbf{0.330} & 0.415 & \textbf{0.352} & 0.577 & \textbf{0.588} & 0.241 & \textbf{0.929} & 0.230 & 0.579 & 0.990 \\ 
  VAR1 & 0.544 & 0.480 & 0.793 & 0.415 & \textbf{0.120} & 0.498 & 0.366 & 0.496 & 0.372 & 0.639 & 0.880 & 0.333 & 1.250 & \textbf{0.222} & 0.665 & 1.078 \\ 
  GFT & -- & 1.037 & \textbf{0.463} & 1.093 & 0.510 & 2.370 & 0.665 & 0.662 & -- & -- & -- & -- & -- & -- & -- & -- \\ 
  naive & 0.541 & 0.482 & 0.799 & 0.421 & 0.121 & 0.498 & 0.367 & 0.489 & 0.381 & 0.627 & 0.868 & 0.333 & 1.229 & 0.225 & 0.667 & 1.047 \\ 
   \hline \hline \multicolumn{1}{l}{MAE}\\
\hline
\end{tabular}
\caption{Comparison of different methods for regional level \%ILI estimation in Region 2.  The RMSE, MAE, and correlation measures are reported. The method with the best performance is highlighted in boldface for each metric in each period. \label{tab:reg2}} 
\end{adjustwidth}\end{table*}


\begin{table*}\begin{adjustwidth}{0cm}{0cm} \tiny
\centering
\begin{tabular}{c|llllllllllllllll}
  \hline
  & '09-'23 & '09-'15 & '09-'10 & '10-'11 & '11-'12 & '12-'13 & '13-'14 & '14-'15 & '15-'16 & '16-'17 & '17-'18 & '18-'19 & '19-'20 & '20-'21 & '21-'22 & '22-'23 \\ 
  \hline \hline \multicolumn{1}{l}{RMSE}\\ARGO-C & \textbf{0.324} & \textbf{0.393} & 0.755 & \textbf{0.264} & \textbf{0.175} & \textbf{0.597} & \textbf{0.230} & 0.458 & \textbf{0.219} & \textbf{0.367} & \textbf{0.379} & \textbf{0.194} & \textbf{0.295} & 0.128 & \textbf{0.303} & \textbf{0.500} \\ 
  ARGO2 & 0.350 & 0.425 & 0.799 & 0.291 & 0.181 & 0.667 & 0.243 & 0.539 & 0.232 & 0.403 & 0.381 & 0.217 & 0.342 & 0.125 & 0.335 & 0.564 \\ 
  VAR1 & 0.460 & 0.522 & 0.998 & 0.327 & 0.193 & 0.786 & 0.311 & 0.745 & 0.320 & 0.530 & 0.652 & 0.334 & 0.708 & \textbf{0.104} & 0.450 & 0.698 \\ 
  GFT & -- & 0.859 & \textbf{0.604} & 0.501 & 0.926 & 2.188 & 0.514 & \textbf{0.380} & -- & -- & -- & -- & -- & -- & -- & -- \\ 
  naive & 0.457 & 0.522 & 0.993 & 0.337 & 0.196 & 0.785 & 0.312 & 0.745 & 0.323 & 0.506 & 0.645 & 0.329 & 0.699 & 0.106 & 0.454 & 0.673 \\ 
   \hline \hline \multicolumn{1}{l}{MAE}\\
\hline
\end{tabular}
\caption{Comparison of different methods for regional level \%ILI estimation in Region 3.  The RMSE, MAE, and correlation measures are reported. The method with the best performance is highlighted in boldface for each metric in each period. \label{tab:reg3}} 
\end{adjustwidth}\end{table*}
\begin{table*}\begin{adjustwidth}{0cm}{0cm} \tiny
\centering
\begin{tabular}{c|llllllllllllllll}
  \hline
  & '09-'23 & '09-'15 & '09-'10 & '10-'11 & '11-'12 & '12-'13 & '13-'14 & '14-'15 & '15-'16 & '16-'17 & '17-'18 & '18-'19 & '19-'20 & '20-'21 & '21-'22 & '22-'23 \\ 
  \hline \hline \multicolumn{1}{l}{RMSE}\\ARGO-C & \textbf{0.335} & \textbf{0.356} & 0.313 & \textbf{0.377} & \textbf{0.180} & 0.524 & \textbf{0.243} & 0.476 & \textbf{0.179} & \textbf{0.390} & 0.421 & \textbf{0.239} & \textbf{0.531} & 0.139 & \textbf{0.286} & \textbf{0.726} \\ 
  ARGO2 & 0.353 & 0.357 & \textbf{0.276} & 0.394 & 0.192 & \textbf{0.524} & 0.267 & 0.517 & 0.210 & 0.415 & \textbf{0.418} & 0.300 & 0.662 & 0.141 & 0.321 & 0.776 \\ 
  VAR1 & 0.463 & 0.422 & 0.286 & 0.477 & 0.203 & 0.568 & 0.386 & 0.728 & 0.281 & 0.505 & 0.817 & 0.483 & 0.923 & \textbf{0.098} & 0.466 & 0.994 \\ 
  GFT & -- & 0.821 & 0.522 & 0.908 & 0.270 & 2.023 & 0.407 & \textbf{0.358} & -- & -- & -- & -- & -- & -- & -- & -- \\ 
  naive & 0.460 & 0.423 & 0.312 & 0.484 & 0.205 & 0.569 & 0.387 & 0.734 & 0.275 & 0.482 & 0.809 & 0.492 & 0.916 & 0.100 & 0.465 & 0.958 \\ 
   \hline \hline \multicolumn{1}{l}{MAE}\\
\hline
\end{tabular}
\caption{Comparison of different methods for regional level \%ILI estimation in Region 4.  The RMSE, MAE, and correlation measures are reported. The method with the best performance is highlighted in boldface for each metric in each period. \label{tab:reg4}} 
\end{adjustwidth}\end{table*}
\begin{table*}\begin{adjustwidth}{0cm}{0cm} \tiny
\centering
\begin{tabular}{c|llllllllllllllll}
  \hline
  & '09-'23 & '09-'15 & '09-'10 & '10-'11 & '11-'12 & '12-'13 & '13-'14 & '14-'15 & '15-'16 & '16-'17 & '17-'18 & '18-'19 & '19-'20 & '20-'21 & '21-'22 & '22-'23 \\ 
  \hline \hline \multicolumn{1}{l}{RMSE}\\ARGO-C & \textbf{0.275} & \textbf{0.326} & 0.729 & \textbf{0.241} & \textbf{0.162} & \textbf{0.378} & \textbf{0.197} & 0.388 & \textbf{0.161} & \textbf{0.267} & \textbf{0.201} & 0.233 & \textbf{0.381} & 0.094 & \textbf{0.247} & \textbf{0.534} \\ 
  ARGO2 & 0.296 & 0.347 & 0.776 & 0.262 & 0.176 & 0.404 & 0.220 & 0.425 & 0.179 & 0.283 & 0.219 & \textbf{0.233} & 0.448 & 0.099 & 0.259 & 0.605 \\ 
  VAR1 & 0.381 & 0.431 & 0.932 & 0.304 & 0.202 & 0.491 & 0.319 & 0.614 & 0.271 & 0.356 & 0.482 & 0.301 & 0.603 & \textbf{0.066} & 0.320 & 0.765 \\ 
  GFT & -- & 0.803 & \textbf{0.480} & 0.382 & 0.528 & 2.311 & 0.323 & \textbf{0.266} & -- & -- & -- & -- & -- & -- & -- & -- \\ 
  naive & 0.378 & 0.430 & 0.917 & 0.308 & 0.202 & 0.487 & 0.321 & 0.630 & 0.271 & 0.343 & 0.480 & 0.301 & 0.596 & 0.067 & 0.317 & 0.738 \\ 
   \hline \hline \multicolumn{1}{l}{MAE}\\
\hline
\end{tabular}
\caption{Comparison of different methods for regional level \%ILI estimation in Region 5.  The RMSE, MAE, and correlation measures are reported. The method with the best performance is highlighted in boldface for each metric in each period. \label{tab:reg5}} 
\end{adjustwidth}\end{table*}
\begin{table*}\begin{adjustwidth}{0cm}{0cm} \tiny
\centering
\begin{tabular}{c|llllllllllllllll}
  \hline
  & '09-'23 & '09-'15 & '09-'10 & '10-'11 & '11-'12 & '12-'13 & '13-'14 & '14-'15 & '15-'16 & '16-'17 & '17-'18 & '18-'19 & '19-'20 & '20-'21 & '21-'22 & '22-'23 \\ 
  \hline \hline \multicolumn{1}{l}{RMSE}\\ARGO-C & \textbf{0.557} & \textbf{0.615} & 0.893 & \textbf{0.466} & \textbf{0.239} & \textbf{0.541} & \textbf{0.548} & 1.117 & 0.362 & \textbf{0.593} & 0.803 & \textbf{0.517} & \textbf{1.017} & 0.298 & 0.311 & \textbf{0.625} \\ 
  ARGO2 & 0.578 & 0.631 & \textbf{0.884} & 0.507 & 0.273 & 0.612 & 0.572 & \textbf{1.108} & \textbf{0.362} & 0.647 & \textbf{0.763} & 0.553 & 1.144 & 0.314 & \textbf{0.303} & 0.663 \\ 
  VAR1 & 0.685 & 0.713 & 0.977 & 0.664 & 0.309 & 0.759 & 0.759 & 1.160 & 0.441 & 0.762 & 1.143 & 0.795 & 1.315 & \textbf{0.249} & 0.451 & 0.785 \\ 
  GFT & -- & 1.356 & 0.899 & 0.639 & 0.908 & 3.607 & 0.770 & 1.146 & -- & -- & -- & -- & -- & -- & -- & -- \\ 
  naive & 0.678 & 0.708 & 1.012 & 0.665 & 0.306 & 0.770 & 0.747 & 1.164 & 0.444 & 0.727 & 1.131 & 0.791 & 1.300 & 0.251 & 0.456 & 0.753 \\ 
   \hline \hline \multicolumn{1}{l}{MAE}\\
\hline
\end{tabular}
\caption{Comparison of different methods for regional level \%ILI estimation in Region 6.  The RMSE, MAE, and correlation measures are reported. The method with the best performance is highlighted in boldface for each metric in each period. \label{tab:reg6}} 
\end{adjustwidth}\end{table*}
\begin{table*}\begin{adjustwidth}{0cm}{0cm} \tiny
\centering
\begin{tabular}{c|llllllllllllllll}
  \hline
  & '09-'23 & '09-'15 & '09-'10 & '10-'11 & '11-'12 & '12-'13 & '13-'14 & '14-'15 & '15-'16 & '16-'17 & '17-'18 & '18-'19 & '19-'20 & '20-'21 & '21-'22 & '22-'23 \\ 
  \hline \hline \multicolumn{1}{l}{RMSE}\\ARGO-C & \textbf{0.423} & \textbf{0.463} & \textbf{0.921} & \textbf{0.387} & \textbf{0.298} & \textbf{0.470} & \textbf{0.263} & \textbf{0.377} & \textbf{0.211} & \textbf{0.448} & \textbf{0.461} & \textbf{0.487} & \textbf{0.584} & 0.260 & 0.268 & \textbf{0.781} \\ 
  ARGO2 & 0.453 & 0.493 & 0.969 & 0.402 & 0.324 & 0.535 & 0.309 & 0.459 & 0.212 & 0.489 & 0.510 & 0.515 & 0.709 & 0.266 & \textbf{0.255} & 0.840 \\ 
  VAR1 & 0.552 & 0.572 & 1.122 & 0.491 & 0.348 & 0.631 & 0.459 & 0.621 & 0.245 & 0.583 & 0.908 & 0.656 & 0.917 & \textbf{0.231} & 0.305 & 1.006 \\ 
  GFT & -- & 0.917 & 1.968 & 1.077 & 0.603 & 0.718 & 0.322 & 0.856 & -- & -- & -- & -- & -- & -- & -- & -- \\ 
  naive & 0.546 & 0.569 & 1.111 & 0.495 & 0.343 & 0.637 & 0.468 & 0.623 & 0.246 & 0.555 & 0.903 & 0.651 & 0.903 & 0.236 & 0.312 & 0.964 \\ 
   \hline \hline \multicolumn{1}{l}{MAE}\\
\hline
\end{tabular}
\caption{Comparison of different methods for regional level \%ILI estimation in Region 7.  The RMSE, MAE, and correlation measures are reported. The method with the best performance is highlighted in boldface for each metric in each period. \label{tab:reg7}} 
\end{adjustwidth}\end{table*}
\begin{table*}\begin{adjustwidth}{0cm}{0cm} \tiny
\centering
\begin{tabular}{c|llllllllllllllll}
  \hline
  & '09-'23 & '09-'15 & '09-'10 & '10-'11 & '11-'12 & '12-'13 & '13-'14 & '14-'15 & '15-'16 & '16-'17 & '17-'18 & '18-'19 & '19-'20 & '20-'21 & '21-'22 & '22-'23 \\ 
  \hline \hline \multicolumn{1}{l}{RMSE}\\ARGO-C & \textbf{0.307} & \textbf{0.336} & \textbf{0.812} & 0.237 & \textbf{0.152} & \textbf{0.213} & 0.283 & \textbf{0.294} & \textbf{0.132} & 0.351 & \textbf{0.169} & \textbf{0.470} & \textbf{0.361} & 0.134 & \textbf{0.244} & \textbf{0.710} \\ 
  ARGO2 & 0.319 & 0.347 & 0.831 & \textbf{0.219} & 0.169 & 0.252 & 0.300 & 0.323 & 0.151 & \textbf{0.337} & 0.172 & 0.476 & 0.415 & 0.139 & 0.259 & 0.758 \\ 
  VAR1 & 0.392 & 0.410 & 0.985 & 0.226 & 0.180 & 0.341 & 0.369 & 0.389 & 0.189 & 0.363 & 0.260 & 0.562 & 0.727 & \textbf{0.088} & 0.317 & 0.919 \\ 
  GFT & -- & 0.622 & 1.502 & 0.382 & 0.387 & 0.912 & \textbf{0.282} & 0.455 & -- & -- & -- & -- & -- & -- & -- & -- \\ 
  naive & 0.385 & 0.402 & 0.969 & 0.229 & 0.177 & 0.341 & 0.374 & 0.388 & 0.186 & 0.369 & 0.250 & 0.532 & 0.721 & 0.088 & 0.315 & 0.894 \\ 
   \hline \hline \multicolumn{1}{l}{MAE}\\
\hline
\end{tabular}
\caption{Comparison of different methods for regional level \%ILI estimation in Region 8.  The RMSE, MAE, and correlation measures are reported. The method with the best performance is highlighted in boldface for each metric in each period. \label{tab:reg8}} 
\end{adjustwidth}\end{table*}
\begin{table*}\begin{adjustwidth}{0cm}{0cm} \tiny
\centering
\begin{tabular}{c|llllllllllllllll}
  \hline
  & '09-'23 & '09-'15 & '09-'10 & '10-'11 & '11-'12 & '12-'13 & '13-'14 & '14-'15 & '15-'16 & '16-'17 & '17-'18 & '18-'19 & '19-'20 & '20-'21 & '21-'22 & '22-'23 \\ 
  \hline \hline \multicolumn{1}{l}{RMSE}\\ARGO-C & \textbf{0.347} & \textbf{0.437} & 0.665 & 0.358 & \textbf{0.382} & \textbf{0.474} & \textbf{0.297} & \textbf{0.284} & \textbf{0.224} & 0.199 & \textbf{0.404} & 0.181 & 0.306 & 0.135 & \textbf{0.229} & \textbf{0.639} \\ 
  ARGO2 & 0.359 & 0.453 & \textbf{0.654} & \textbf{0.353} & 0.394 & 0.525 & 0.319 & 0.330 & 0.244 & \textbf{0.186} & 0.413 & \textbf{0.178} & \textbf{0.288} & \textbf{0.134} & 0.250 & 0.700 \\ 
  VAR1 & 0.406 & 0.470 & 0.664 & 0.395 & 0.419 & 0.538 & 0.382 & 0.380 & 0.315 & 0.208 & 0.622 & 0.224 & 0.462 & 0.137 & 0.353 & 0.877 \\ 
  GFT & -- & 1.091 & 1.101 & 1.168 & 0.567 & 2.536 & 0.593 & 0.540 & -- & -- & -- & -- & -- & -- & -- & -- \\ 
  naive & 0.405 & 0.475 & 0.676 & 0.382 & 0.438 & 0.544 & 0.385 & 0.382 & 0.312 & 0.210 & 0.606 & 0.222 & 0.461 & 0.139 & 0.358 & 0.828 \\ 
   \hline \hline \multicolumn{1}{l}{MAE}\\
\hline
\end{tabular}
\caption{Comparison of different methods for regional level \%ILI estimation in Region 9.  The RMSE, MAE, and correlation measures are reported.. The method with the best performance is highlighted in boldface for each metric in each period. \label{tab:reg9}} 
\end{adjustwidth}\end{table*}
\begin{table*}\begin{adjustwidth}{0cm}{0cm} \tiny
\centering
\begin{tabular}{c|llllllllllllllll}
  \hline
  & '09-'23 & '09-'15 & '09-'10 & '10-'11 & '11-'12 & '12-'13 & '13-'14 & '14-'15 & '15-'16 & '16-'17 & '17-'18 & '18-'19 & '19-'20 & '20-'21 & '21-'22 & '22-'23 \\ 
  \hline \hline \multicolumn{1}{l}{RMSE}\\ARGO-C & \textbf{0.383} & \textbf{0.419} & 0.697 & 0.466 & 0.225 & \textbf{0.379} & \textbf{0.376} & \textbf{0.259} & \textbf{0.254} & \textbf{0.438} & \textbf{0.345} & \textbf{0.282} & \textbf{0.453} & 0.122 & 0.205 & \textbf{1.144} \\ 
  ARGO2 & 0.396 & 0.428 & \textbf{0.687} & 0.469 & \textbf{0.221} & 0.406 & 0.399 & 0.267 & 0.259 & 0.452 & 0.350 & 0.308 & 0.530 & 0.120 & \textbf{0.179} & 1.194 \\ 
  VAR1 & 0.471 & 0.469 & 0.733 & 0.477 & 0.235 & 0.438 & 0.463 & 0.330 & 0.277 & 0.493 & 0.504 & 0.410 & 0.774 & \textbf{0.085} & 0.267 & 1.526 \\ 
  GFT & -- & 1.083 & 0.800 & \textbf{0.423} & 0.629 & 2.880 & 0.869 & 0.339 & -- & -- & -- & -- & -- & -- & -- & -- \\ 
  naive & 0.457 & 0.457 & 0.705 & 0.493 & 0.237 & 0.451 & 0.467 & 0.325 & 0.271 & 0.506 & 0.505 & 0.400 & 0.783 & 0.085 & 0.259 & 1.393 \\ 
   \hline \hline \multicolumn{1}{l}{MAE}\\
\hline
\end{tabular}
\caption{Comparison of different methods for regional level \%ILI estimation in Region 10.  The RMSE, MAE, and correlation measures are reported. The method with the best performance is highlighted in boldface for each metric in each period. \label{tab:reg10}} 
\end{adjustwidth}\end{table*}

%% file: tables_state_error.tex
\begin{table*}
\centering
\begingroup\scriptsize
\begin{tabular}{c|lllllllllllll}
  \hline
  & Whole period & '14-'23 & post-COVID & GFT period & '14-'15 & '15-'16 & '16-'17 & '17-'18 & '18-'19 & '19-'20 & '20-'21 & '21-'22 & '22-'23 \\ 
  \hline \hline \multicolumn{1}{l}{RMSE}\\ARGO-C & \textbf{0.831} & \textbf{0.752} & \textbf{0.577} & \textbf{1.756} & \textbf{1.922} & 0.625 & \textbf{0.544} & \textbf{0.846} & \textbf{0.504} & \textbf{0.698} & 0.198 & \textbf{0.274} & \textbf{1.507} \\ 
  ARGOX & 0.873 & 0.797 & 0.632 & 1.865 & 2.045 & 0.607 & 0.596 & 0.864 & 0.530 & 0.779 & 0.192 & 0.289 & 1.653 \\ 
  VAR1 & 2.407 & 2.383 & 2.336 & 4.624 & 5.083 & 1.324 & 2.231 & 3.394 & 1.693 & 3.559 & 0.351 & 0.687 & 5.138 \\ 
  GFT & -- & -- & -- & 2.891 & 3.183 & -- & -- & -- & -- & -- & -- & -- & -- \\ 
  naive & 1.115 & 0.977 & 0.646 & 2.271 & 2.495 & \textbf{0.591} & 0.831 & 1.315 & 0.785 & 1.043 & \textbf{0.142} & 0.453 & 1.645 \\ 
   \hline \hline \multicolumn{1}{l}{MAE}\\
\hline
\end{tabular}
\endgroup
\caption{Comparison of different methods for state-level \%ILI estimation in Alabama.  The MSE, MAE, and correlation are reported. The method with the best performance is highlighted in boldface for each metric in each period. \label{tab:state US.AL}} 
\end{table*}
\begin{table*}
\centering
\begingroup\scriptsize
\begin{tabular}{c|lllllllllllll}
  \hline
  & Whole period & '14-'23 & post-COVID & GFT period & '14-'15 & '15-'16 & '16-'17 & '17-'18 & '18-'19 & '19-'20 & '20-'21 & '21-'22 & '22-'23 \\ 
  \hline \hline \multicolumn{1}{l}{RMSE}\\ARGO-C & \textbf{0.961} & \textbf{0.849} & \textbf{0.586} & 0.663 & 0.630 & \textbf{0.828} & \textbf{0.705} & \textbf{0.634} & \textbf{1.097} & 1.962 & 0.301 & \textbf{0.631} & \textbf{0.625} \\ 
  ARGOX & 0.966 & 0.855 & 0.596 & 0.654 & 0.622 & 0.832 & 0.707 & 0.635 & 1.108 & 1.979 & 0.298 & 0.655 & 0.630 \\ 
  VAR1 & 1.058 & 0.945 & 0.690 & 0.689 & 0.656 & 1.159 & 0.929 & 1.095 & 1.215 & \textbf{1.847} & 0.305 & 0.789 & 1.071 \\ 
  GFT & -- & -- & -- & 1.015 & 1.042 & -- & -- & -- & -- & -- & -- & -- & -- \\ 
  naive & 1.002 & 0.896 & 0.654 & \textbf{0.646} & \textbf{0.605} & 0.843 & 0.783 & 0.645 & 1.126 & 2.047 & \textbf{0.287} & 0.788 & 0.660 \\ 
   \hline \hline \multicolumn{1}{l}{MAE}\\
\hline
\end{tabular}
\endgroup
\caption{Comparison of different methods for state-level \%ILI estimation in Alaska.  The MSE, MAE, and correlation are reported. The method with the best performance is highlighted in boldface for each metric in each period. \label{tab:state US.AK}} 
\end{table*}
\begin{table*}
\centering
\begingroup\scriptsize
\begin{tabular}{c|lllllllllllll}
  \hline
  & Whole period & '14-'23 & post-COVID & GFT period & '14-'15 & '15-'16 & '16-'17 & '17-'18 & '18-'19 & '19-'20 & '20-'21 & '21-'22 & '22-'23 \\ 
  \hline \hline \multicolumn{1}{l}{RMSE}\\ARGO-C & \textbf{0.396} & \textbf{0.378} & 0.342 & \textbf{0.246} & \textbf{0.256} & \textbf{0.391} & 0.319 & \textbf{0.587} & 0.315 & 0.662 & 0.272 & 0.336 & \textbf{0.483} \\ 
  ARGOX & 0.404 & 0.382 & \textbf{0.338} & 0.253 & 0.266 & 0.413 & 0.311 & 0.645 & \textbf{0.309} & \textbf{0.629} & \textbf{0.259} & \textbf{0.335} & 0.513 \\ 
  VAR1 & 2.372 & 2.080 & 1.384 & 4.487 & 4.938 & 1.218 & 2.423 & 2.891 & 1.615 & 2.398 & 0.365 & 0.539 & 3.543 \\ 
  GFT & -- & -- & -- & 1.160 & 1.253 & -- & -- & -- & -- & -- & -- & -- & -- \\ 
  naive & 0.454 & 0.432 & 0.387 & 0.280 & 0.298 & 0.401 & \textbf{0.304} & 0.787 & 0.363 & 0.707 & 0.279 & 0.424 & 0.679 \\ 
   \hline \hline \multicolumn{1}{l}{MAE}\\
\hline
\end{tabular}
\endgroup
\caption{Comparison of different methods for state-level \%ILI estimation in Arizona.  The MSE, MAE, and correlation are reported. The method with the best performance is highlighted in boldface for each metric in each period. \label{tab:state US.AZ}} 
\end{table*}
\begin{table*}
\centering
\begingroup\scriptsize
\begin{tabular}{c|lllllllllllll}
  \hline
  & Whole period & '14-'23 & post-COVID & GFT period & '14-'15 & '15-'16 & '16-'17 & '17-'18 & '18-'19 & '19-'20 & '20-'21 & '21-'22 & '22-'23 \\ 
  \hline \hline \multicolumn{1}{l}{RMSE}\\ARGO-C & \textbf{0.821} & \textbf{0.741} & \textbf{0.564} & \textbf{1.050} & \textbf{1.144} & 0.622 & \textbf{0.842} & \textbf{0.905} & 0.830 & \textbf{1.576} & 0.195 & 0.405 & \textbf{0.718} \\ 
  ARGOX & 0.853 & 0.780 & 0.622 & 1.102 & 1.201 & \textbf{0.612} & 0.897 & 0.973 & \textbf{0.805} & 1.716 & 0.199 & \textbf{0.382} & 0.798 \\ 
  VAR1 & 1.103 & 0.950 & 0.565 & 1.382 & 1.514 & 0.656 & 1.154 & 1.597 & 1.041 & 1.812 & \textbf{0.167} & 0.417 & 0.885 \\ 
  GFT & -- & -- & -- & 1.391 & 1.466 & -- & -- & -- & -- & -- & -- & -- & -- \\ 
  naive & 1.049 & 0.919 & 0.609 & 1.328 & 1.453 & 0.647 & 1.080 & 1.444 & 0.987 & 1.871 & 0.168 & 0.421 & 0.846 \\ 
   \hline \hline \multicolumn{1}{l}{MAE}\\
\hline
\end{tabular}
\endgroup
\caption{Comparison of different methods for state-level \%ILI estimation in Arkansas.  The MSE, MAE, and correlation are reported. The method with the best performance is highlighted in boldface for each metric in each period. \label{tab:state US.AR}} 
\end{table*}
\begin{table*}
\centering
\begingroup\scriptsize
\begin{tabular}{c|lllllllllllll}
  \hline
  & Whole period & '14-'23 & post-COVID & GFT period & '14-'15 & '15-'16 & '16-'17 & '17-'18 & '18-'19 & '19-'20 & '20-'21 & '21-'22 & '22-'23 \\ 
  \hline \hline \multicolumn{1}{l}{RMSE}\\ARGO-C & \textbf{0.252} & \textbf{0.267} & \textbf{0.292} & \textbf{0.255} & \textbf{0.275} & \textbf{0.278} & 0.247 & \textbf{0.389} & 0.235 & 0.342 & 0.142 & \textbf{0.248} & \textbf{0.649} \\ 
  ARGOX & 0.260 & 0.279 & 0.312 & 0.276 & 0.298 & 0.303 & \textbf{0.243} & 0.416 & \textbf{0.217} & \textbf{0.326} & \textbf{0.133} & 0.261 & 0.738 \\ 
  VAR1 & 0.793 & 0.872 & 1.003 & 1.287 & 1.347 & 0.648 & 0.640 & 0.987 & 0.595 & 0.810 & 0.264 & 0.671 & 2.477 \\ 
  GFT & -- & -- & -- & 0.535 & 0.555 & -- & -- & -- & -- & -- & -- & -- & -- \\ 
  naive & 0.344 & 0.372 & 0.419 & 0.370 & 0.402 & 0.354 & 0.252 & 0.640 & 0.262 & 0.492 & 0.150 & 0.410 & 0.937 \\ 
   \hline \hline \multicolumn{1}{l}{MAE}\\
\hline
\end{tabular}
\endgroup
\caption{Comparison of different methods for state-level \%ILI estimation in California.  The MSE, MAE, and correlation are reported. The method with the best performance is highlighted in boldface for each metric in each period. \label{tab:state US.CA}} 
\end{table*}
\begin{table*}
\centering
\begingroup\scriptsize
\begin{tabular}{c|lllllllllllll}
  \hline
  & Whole period & '14-'23 & post-COVID & GFT period & '14-'15 & '15-'16 & '16-'17 & '17-'18 & '18-'19 & '19-'20 & '20-'21 & '21-'22 & '22-'23 \\ 
  \hline \hline \multicolumn{1}{l}{RMSE}\\ARGO-C & 0.422 & \textbf{0.410} & \textbf{0.386} & 0.410 & 0.450 & \textbf{0.151} & 0.608 & 0.225 & 0.790 & \textbf{0.422} & 0.150 & \textbf{0.370} & \textbf{0.862} \\ 
  ARGOX & \textbf{0.417} & 0.411 & 0.401 & \textbf{0.402} & \textbf{0.441} & 0.155 & \textbf{0.589} & \textbf{0.207} & 0.783 & 0.443 & 0.146 & 0.384 & 0.906 \\ 
  VAR1 & 1.414 & 1.289 & 1.019 & 1.671 & 1.669 & 1.512 & 0.880 & 1.602 & 1.921 & 1.829 & 0.621 & 1.340 & 1.582 \\ 
  GFT & -- & -- & -- & 0.809 & 0.886 & -- & -- & -- & -- & -- & -- & -- & -- \\ 
  naive & 0.461 & 0.489 & 0.536 & 0.478 & 0.525 & 0.208 & 0.685 & 0.270 & \textbf{0.725} & 0.829 & \textbf{0.130} & 0.376 & 1.143 \\ 
   \hline \hline \multicolumn{1}{l}{MAE}\\
\hline
\end{tabular}
\endgroup
\caption{Comparison of different methods for state-level \%ILI estimation in Colorado.  The MSE, MAE, and correlation are reported. The method with the best performance is highlighted in boldface for each metric in each period. \label{tab:state US.CO}} 
\end{table*}
\begin{table*}
\centering
\begingroup\scriptsize
\begin{tabular}{c|lllllllllllll}
  \hline
  & Whole period & '14-'23 & post-COVID & GFT period & '14-'15 & '15-'16 & '16-'17 & '17-'18 & '18-'19 & '19-'20 & '20-'21 & '21-'22 & '22-'23 \\ 
  \hline \hline \multicolumn{1}{l}{RMSE}\\ARGO-C & \textbf{0.534} & \textbf{0.489} & \textbf{0.393} & 0.677 & 0.743 & 0.471 & 0.409 & 0.462 & 0.445 & \textbf{0.724} & 0.162 & \textbf{0.337} & \textbf{0.734} \\ 
  ARGOX & 0.536 & 0.493 & 0.400 & 0.680 & 0.746 & 0.471 & \textbf{0.395} & \textbf{0.447} & \textbf{0.412} & 0.769 & 0.158 & 0.371 & 0.762 \\ 
  VAR1 & 1.425 & 1.274 & 0.933 & 1.684 & 1.840 & 1.548 & 1.589 & 2.358 & 0.828 & 1.163 & 0.267 & 1.012 & 2.018 \\ 
  GFT & -- & -- & -- & 1.475 & 1.614 & -- & -- & -- & -- & -- & -- & -- & -- \\ 
  naive & 0.576 & 0.537 & 0.456 & \textbf{0.625} & \textbf{0.683} & \textbf{0.462} & 0.492 & 0.575 & 0.493 & 0.865 & \textbf{0.149} & 0.414 & 0.935 \\ 
   \hline \hline \multicolumn{1}{l}{MAE}\\
\hline
\end{tabular}
\endgroup
\caption{Comparison of different methods for state-level \%ILI estimation in Connecticut.  The MSE, MAE, and correlation are reported. The method with the best performance is highlighted in boldface for each metric in each period. \label{tab:state US.CT}} 
\end{table*}
\begin{table*}
\centering
\begingroup\scriptsize
\begin{tabular}{c|lllllllllllll}
  \hline
  & Whole period & '14-'23 & post-COVID & GFT period & '14-'15 & '15-'16 & '16-'17 & '17-'18 & '18-'19 & '19-'20 & '20-'21 & '21-'22 & '22-'23 \\ 
  \hline \hline \multicolumn{1}{l}{RMSE}\\ARGO-C & \textbf{0.421} & \textbf{0.404} & \textbf{0.370} & 0.549 & 0.604 & 0.252 & 0.203 & 0.689 & 0.256 & \textbf{0.808} & 0.190 & \textbf{0.293} & \textbf{0.315} \\ 
  ARGOX & 0.421 & 0.405 & 0.373 & 0.553 & 0.608 & \textbf{0.243} & \textbf{0.186} & \textbf{0.684} & \textbf{0.254} & 0.836 & \textbf{0.186} & 0.295 & 0.324 \\ 
  VAR1 & 1.898 & 1.755 & 1.453 & 1.747 & 1.882 & 1.808 & 2.227 & 2.454 & 2.420 & 3.279 & 0.426 & 0.696 & 2.519 \\ 
  GFT & -- & -- & -- & 2.085 & 2.232 & -- & -- & -- & -- & -- & -- & -- & -- \\ 
  naive & 0.453 & 0.453 & 0.452 & \textbf{0.509} & \textbf{0.560} & 0.256 & 0.230 & 0.785 & 0.293 & 0.954 & 0.202 & 0.376 & 0.502 \\ 
   \hline \hline \multicolumn{1}{l}{MAE}\\
\hline
\end{tabular}
\endgroup
\caption{Comparison of different methods for state-level \%ILI estimation in Delaware.  The MSE, MAE, and correlation are reported. The method with the best performance is highlighted in boldface for each metric in each period. \label{tab:state US.DE}} 
\end{table*}
\begin{table*}
\centering
\begingroup\scriptsize
\begin{tabular}{c|lllllllllllll}
  \hline
  & Whole period & '14-'23 & post-COVID & GFT period & '14-'15 & '15-'16 & '16-'17 & '17-'18 & '18-'19 & '19-'20 & '20-'21 & '21-'22 & '22-'23 \\ 
  \hline \hline \multicolumn{1}{l}{RMSE}\\ARGO-C & 1.421 & 1.264 & 0.904 & 1.515 & 1.596 & \textbf{1.595} & \textbf{2.054} & 1.190 & 0.494 & 0.802 & 0.508 & 0.728 & 1.660 \\ 
  ARGOX & \textbf{1.418} & \textbf{1.262} & 0.902 & \textbf{1.509} & \textbf{1.587} & 1.615 & 2.062 & 1.226 & \textbf{0.459} & 0.798 & 0.512 & \textbf{0.714} & 1.673 \\ 
  VAR1 & 1.523 & 1.346 & 0.934 & 1.560 & 1.651 & 1.632 & 2.157 & 1.604 & 0.542 & 0.861 & \textbf{0.459} & 0.777 & 1.601 \\ 
  GFT & -- & -- & -- & 5.236 & 5.015 & -- & -- & -- & -- & -- & -- & -- & -- \\ 
  naive & 1.447 & 1.276 & \textbf{0.872} & 1.588 & 1.676 & 1.669 & 2.193 & \textbf{0.846} & 0.529 & \textbf{0.770} & 0.496 & 0.796 & \textbf{1.465} \\ 
   \hline \hline \multicolumn{1}{l}{MAE}\\
\hline
\end{tabular}
\endgroup
\caption{Comparison of different methods for state-level \%ILI estimation in District of Columbia.  The MSE, MAE, and correlation are reported. The method with the best performance is highlighted in boldface for each metric in each period. \label{tab:state US.DC}} 
\end{table*}
\begin{table*}
\centering
\begingroup\scriptsize
\begin{tabular}{c|lllllllllllll}
  \hline
  & Whole period & '14-'23 & post-COVID & GFT period & '14-'15 & '15-'16 & '16-'17 & '17-'18 & '18-'19 & '19-'20 & '20-'21 & '21-'22 & '22-'23 \\ 
  \hline \hline \multicolumn{1}{l}{RMSE}\\ARGO-C & \textbf{0.577} & \textbf{0.543} & \textbf{0.473} & 0.860 & 0.945 & \textbf{0.318} & \textbf{0.421} & 1.031 & \textbf{0.401} & \textbf{0.700} & 0.247 & \textbf{0.446} & \textbf{0.856} \\ 
  ARGOX & 0.593 & 0.560 & 0.492 & 0.871 & 0.957 & 0.330 & 0.437 & \textbf{1.017} & 0.459 & 0.753 & 0.249 & 0.482 & 0.951 \\ 
  VAR1 & 3.205 & 2.992 & 2.550 & 2.220 & 2.421 & 1.629 & 3.280 & 5.718 & 3.521 & 4.928 & 1.583 & 2.289 & 4.099 \\ 
  GFT & -- & -- & -- & \textbf{0.819} & \textbf{0.896} & -- & -- & -- & -- & -- & -- & -- & -- \\ 
  naive & 0.816 & 0.758 & 0.638 & 0.999 & 1.098 & 0.358 & 0.564 & 1.501 & 0.672 & 1.360 & \textbf{0.206} & 0.740 & 0.935 \\ 
   \hline \hline \multicolumn{1}{l}{MAE}\\
\hline
\end{tabular}
\endgroup
\caption{Comparison of different methods for state-level \%ILI estimation in Georgia.  The MSE, MAE, and correlation are reported. The method with the best performance is highlighted in boldface for each metric in each period. \label{tab:state US.GA}} 
\end{table*}
\begin{table*}
\centering
\begingroup\scriptsize
\begin{tabular}{c|lllllllllllll}
  \hline
  & Whole period & '14-'23 & post-COVID & GFT period & '14-'15 & '15-'16 & '16-'17 & '17-'18 & '18-'19 & '19-'20 & '20-'21 & '21-'22 & '22-'23 \\ 
  \hline \hline \multicolumn{1}{l}{RMSE}\\ARGO-C & \textbf{0.903} & \textbf{0.775} & 0.450 & 1.659 & 1.702 & \textbf{0.769} & 0.688 & \textbf{1.155} & \textbf{0.611} & 0.648 & 0.354 & 0.349 & \textbf{0.471} \\ 
  ARGOX & 0.906 & 0.778 & 0.453 & 1.660 & 1.704 & 0.773 & \textbf{0.684} & 1.164 & 0.625 & 0.643 & \textbf{0.350} & 0.356 & 0.485 \\ 
  VAR1 & 2.945 & 2.647 & 1.978 & 5.999 & 6.525 & 1.870 & 0.973 & 3.918 & 1.713 & 3.541 & 1.481 & 1.985 & 2.306 \\ 
  GFT & -- & -- & -- & 4.131 & 4.457 & -- & -- & -- & -- & -- & -- & -- & -- \\ 
  naive & 0.930 & 0.796 & \textbf{0.449} & \textbf{1.597} & \textbf{1.627} & 0.787 & 0.740 & 1.269 & 0.707 & \textbf{0.618} & 0.374 & \textbf{0.347} & 0.504 \\ 
   \hline \hline \multicolumn{1}{l}{MAE}\\
\hline
\end{tabular}
\endgroup
\caption{Comparison of different methods for state-level \%ILI estimation in Hawaii.  The MSE, MAE, and correlation are reported. The method with the best performance is highlighted in boldface for each metric in each period. \label{tab:state US.HI}} 
\end{table*}
\begin{table*}
\centering
\begingroup\scriptsize
\begin{tabular}{c|lllllllllllll}
  \hline
  & Whole period & '14-'23 & post-COVID & GFT period & '14-'15 & '15-'16 & '16-'17 & '17-'18 & '18-'19 & '19-'20 & '20-'21 & '21-'22 & '22-'23 \\ 
  \hline \hline \multicolumn{1}{l}{RMSE}\\ARGO-C & 0.508 & \textbf{0.622} & \textbf{0.791} & 0.651 & 0.670 & 0.530 & \textbf{0.411} & 0.568 & \textbf{0.730} & 0.906 & 0.233 & 0.578 & \textbf{1.475} \\ 
  ARGOX & \textbf{0.495} & 0.623 & 0.809 & \textbf{0.632} & \textbf{0.651} & 0.499 & 0.420 & \textbf{0.519} & 0.739 & 0.938 & \textbf{0.231} & \textbf{0.569} & 1.515 \\ 
  VAR1 & 2.281 & 2.027 & 1.443 & 4.065 & 4.298 & 2.339 & 1.798 & 2.449 & 1.559 & 1.851 & 0.500 & 1.146 & 3.160 \\ 
  GFT & -- & -- & -- & 1.243 & 1.355 & -- & -- & -- & -- & -- & -- & -- & -- \\ 
  naive & 0.566 & 0.669 & 0.826 & 0.827 & 0.850 & \textbf{0.470} & 0.478 & 0.639 & 0.820 & \textbf{0.808} & 0.234 & 0.595 & 1.672 \\ 
   \hline \hline \multicolumn{1}{l}{MAE}\\
\hline
\end{tabular}
\endgroup
\caption{Comparison of different methods for state-level \%ILI estimation in Idaho.  The MSE, MAE, and correlation are reported. The method with the best performance is highlighted in boldface for each metric in each period. \label{tab:state US.ID}} 
\end{table*}
\begin{table*}
\centering
\begingroup\scriptsize
\begin{tabular}{c|lllllllllllll}
  \hline
  & Whole period & '14-'23 & post-COVID & GFT period & '14-'15 & '15-'16 & '16-'17 & '17-'18 & '18-'19 & '19-'20 & '20-'21 & '21-'22 & '22-'23 \\ 
  \hline \hline \multicolumn{1}{l}{RMSE}\\ARGO-C & \textbf{0.284} & \textbf{0.284} & \textbf{0.284} & 0.460 & 0.503 & 0.233 & \textbf{0.230} & \textbf{0.328} & \textbf{0.269} & \textbf{0.430} & 0.156 & 0.293 & \textbf{0.555} \\ 
  ARGOX & 0.291 & 0.292 & 0.292 & \textbf{0.443} & \textbf{0.485} & \textbf{0.233} & 0.241 & 0.354 & 0.272 & 0.471 & 0.146 & \textbf{0.273} & 0.615 \\ 
  VAR1 & 1.628 & 1.552 & 1.399 & 2.173 & 2.332 & 1.926 & 1.832 & 1.962 & 0.955 & 2.258 & 0.715 & 0.730 & 2.283 \\ 
  GFT & -- & -- & -- & 0.672 & 0.704 & -- & -- & -- & -- & -- & -- & -- & -- \\ 
  naive & 0.401 & 0.396 & 0.387 & 0.459 & 0.501 & 0.271 & 0.360 & 0.626 & 0.393 & 0.765 & \textbf{0.114} & 0.319 & 0.756 \\ 
   \hline \hline \multicolumn{1}{l}{MAE}\\
\hline
\end{tabular}
\endgroup
\caption{Comparison of different methods for state-level \%ILI estimation in Illinois.  The MSE, MAE, and correlation are reported. The method with the best performance is highlighted in boldface for each metric in each period. \label{tab:state US.IL}} 
\end{table*}
\begin{table*}
\centering
\begingroup\scriptsize
\begin{tabular}{c|lllllllllllll}
  \hline
  & Whole period & '14-'23 & post-COVID & GFT period & '14-'15 & '15-'16 & '16-'17 & '17-'18 & '18-'19 & '19-'20 & '20-'21 & '21-'22 & '22-'23 \\ 
  \hline \hline \multicolumn{1}{l}{RMSE}\\ARGO-C & \textbf{0.544} & \textbf{0.512} & \textbf{0.446} & 0.656 & 0.714 & \textbf{0.416} & 0.718 & \textbf{0.571} & \textbf{0.596} & \textbf{0.810} & 0.163 & 0.230 & \textbf{0.690} \\ 
  ARGOX & 0.566 & 0.537 & 0.478 & 0.714 & 0.778 & 0.431 & \textbf{0.713} & 0.640 & 0.613 & 0.884 & 0.165 & \textbf{0.207} & 0.728 \\ 
  VAR1 & 1.770 & 1.584 & 1.162 & 1.788 & 1.936 & 0.968 & 1.866 & 2.985 & 1.888 & 2.675 & 0.172 & 0.633 & 2.147 \\ 
  GFT & -- & -- & -- & \textbf{0.478} & \textbf{0.521} & -- & -- & -- & -- & -- & -- & -- & -- \\ 
  naive & 0.728 & 0.654 & 0.486 & 0.916 & 1.004 & 0.562 & 0.830 & 0.978 & 0.783 & 1.018 & \textbf{0.103} & 0.248 & 0.863 \\ 
   \hline \hline \multicolumn{1}{l}{MAE}\\
\hline
\end{tabular}
\endgroup
\caption{Comparison of different methods for state-level \%ILI estimation in Indiana.  The MSE, MAE, and correlation are reported. The method with the best performance is highlighted in boldface for each metric in each period. \label{tab:state US.IN}} 
\end{table*}
\begin{table*}
\centering
\begingroup\scriptsize
\begin{tabular}{c|lllllllllllll}
  \hline
  & Whole period & '14-'23 & post-COVID & GFT period & '14-'15 & '15-'16 & '16-'17 & '17-'18 & '18-'19 & '19-'20 & '20-'21 & '21-'22 & '22-'23 \\ 
  \hline \hline \multicolumn{1}{l}{RMSE}\\ARGO-C & 0.399 & \textbf{0.686} & \textbf{1.023} & 0.668 & 0.735 & 0.280 & \textbf{0.362} & 0.394 & 0.409 & \textbf{0.537} & 0.930 & \textbf{0.534} & \textbf{1.034} \\ 
  ARGOX & \textbf{0.397} & 0.699 & 1.050 & 0.652 & 0.718 & \textbf{0.267} & 0.388 & \textbf{0.386} & \textbf{0.408} & 0.632 & 0.923 & 0.597 & 1.119 \\ 
  VAR1 & 1.442 & 1.466 & 1.510 & 1.982 & 2.101 & 1.811 & 1.730 & 1.630 & 0.869 & 2.133 & \textbf{0.915} & 0.985 & 1.368 \\ 
  GFT & -- & -- & -- & 2.296 & 2.519 & -- & -- & -- & -- & -- & -- & -- & -- \\ 
  naive & 0.463 & 0.786 & 1.169 & \textbf{0.646} & \textbf{0.711} & 0.294 & 0.425 & 0.554 & 0.478 & 0.749 & 0.995 & 0.608 & 1.386 \\ 
   \hline \hline \multicolumn{1}{l}{MAE}\\
\hline
\end{tabular}
\endgroup
\caption{Comparison of different methods for state-level \%ILI estimation in Iowa.  The MSE, MAE, and correlation are reported. The method with the best performance is highlighted in boldface for each metric in each period. \label{tab:state US.IA}} 
\end{table*}
\begin{table*}
\centering
\begingroup\scriptsize
\begin{tabular}{c|lllllllllllll}
  \hline
  & Whole period & '14-'23 & post-COVID & GFT period & '14-'15 & '15-'16 & '16-'17 & '17-'18 & '18-'19 & '19-'20 & '20-'21 & '21-'22 & '22-'23 \\ 
  \hline \hline \multicolumn{1}{l}{RMSE}\\ARGO-C & \textbf{0.523} & \textbf{0.508} & \textbf{0.478} & \textbf{0.702} & \textbf{0.770} & 0.283 & \textbf{0.864} & 0.533 & 0.507 & \textbf{0.635} & 0.212 & 0.482 & \textbf{0.931} \\ 
  ARGOX & 0.526 & 0.515 & 0.494 & 0.729 & 0.799 & \textbf{0.277} & 0.902 & \textbf{0.481} & \textbf{0.456} & 0.668 & 0.213 & 0.472 & 1.012 \\ 
  VAR1 & 1.444 & 1.251 & 0.774 & 1.783 & 1.958 & 0.470 & 2.102 & 2.230 & 0.877 & 1.951 & 0.249 & 0.614 & 1.280 \\ 
  GFT & -- & -- & -- & 1.472 & 1.620 & -- & -- & -- & -- & -- & -- & -- & -- \\ 
  naive & 0.678 & 0.633 & 0.542 & 0.898 & 0.986 & 0.298 & 0.971 & 1.021 & 0.554 & 0.874 & \textbf{0.196} & \textbf{0.441} & 1.151 \\ 
   \hline \hline \multicolumn{1}{l}{MAE}\\
\hline
\end{tabular}
\endgroup
\caption{Comparison of different methods for state-level \%ILI estimation in Kansas.  The MSE, MAE, and correlation are reported. The method with the best performance is highlighted in boldface for each metric in each period. \label{tab:state US.KS}} 
\end{table*}
\begin{table*}
\centering
\begingroup\scriptsize
\begin{tabular}{c|lllllllllllll}
  \hline
  & Whole period & '14-'23 & post-COVID & GFT period & '14-'15 & '15-'16 & '16-'17 & '17-'18 & '18-'19 & '19-'20 & '20-'21 & '21-'22 & '22-'23 \\ 
  \hline \hline \multicolumn{1}{l}{RMSE}\\ARGO-C & \textbf{0.670} & \textbf{0.631} & \textbf{0.550} & \textbf{0.279} & \textbf{0.306} & 0.254 & \textbf{1.071} & 1.018 & \textbf{0.837} & \textbf{0.997} & 0.215 & 0.391 & \textbf{1.210} \\ 
  ARGOX & 0.683 & 0.656 & 0.603 & 0.293 & 0.322 & \textbf{0.244} & 1.075 & \textbf{0.996} & 0.875 & 1.163 & 0.217 & \textbf{0.347} & 1.280 \\ 
  VAR1 & 1.698 & 1.507 & 1.067 & 2.646 & 2.906 & 0.741 & 1.284 & 1.900 & 2.522 & 2.259 & 0.275 & 0.636 & 2.434 \\ 
  GFT & -- & -- & -- & 2.912 & 3.180 & -- & -- & -- & -- & -- & -- & -- & -- \\ 
  naive & 0.850 & 0.774 & 0.607 & 0.393 & 0.432 & 0.281 & 1.222 & 1.300 & 1.038 & 1.384 & \textbf{0.174} & 0.395 & 1.438 \\ 
   \hline \hline \multicolumn{1}{l}{MAE}\\
\hline
\end{tabular}
\endgroup
\caption{Comparison of different methods for state-level \%ILI estimation in Kentucky.  The MSE, MAE, and correlation are reported. The method with the best performance is highlighted in boldface for each metric in each period. \label{tab:state US.KY}} 
\end{table*}
\begin{table*}
\centering
\begingroup\scriptsize
\begin{tabular}{c|lllllllllllll}
  \hline
  & Whole period & '14-'23 & post-COVID & GFT period & '14-'15 & '15-'16 & '16-'17 & '17-'18 & '18-'19 & '19-'20 & '20-'21 & '21-'22 & '22-'23 \\ 
  \hline \hline \multicolumn{1}{l}{RMSE}\\ARGO-C & \textbf{0.511} & \textbf{0.487} & 0.438 & 0.488 & 0.534 & 0.276 & \textbf{0.274} & \textbf{0.617} & 0.870 & \textbf{0.896} & 0.507 & 0.350 & \textbf{0.569} \\ 
  ARGOX & 0.524 & 0.496 & \textbf{0.438} & \textbf{0.455} & \textbf{0.498} & 0.264 & 0.277 & 0.679 & \textbf{0.866} & 0.937 & 0.491 & \textbf{0.317} & 0.655 \\ 
  VAR1 & 2.297 & 2.099 & 1.671 & 2.946 & 3.199 & 1.567 & 2.045 & 2.380 & 1.876 & 4.363 & 0.345 & 0.769 & 2.875 \\ 
  GFT & -- & -- & -- & 1.015 & 1.106 & -- & -- & -- & -- & -- & -- & -- & -- \\ 
  naive & 0.641 & 0.612 & 0.553 & 0.567 & 0.621 & \textbf{0.249} & 0.355 & 0.917 & 0.981 & 1.373 & \textbf{0.323} & 0.385 & 0.680 \\ 
   \hline \hline \multicolumn{1}{l}{MAE}\\
\hline
\end{tabular}
\endgroup
\caption{Comparison of different methods for state-level \%ILI estimation in Louisiana.  The MSE, MAE, and correlation are reported. The method with the best performance is highlighted in boldface for each metric in each period. \label{tab:state US.LA}} 
\end{table*}
\begin{table*}
\centering
\begingroup\scriptsize
\begin{tabular}{c|lllllllllllll}
  \hline
  & Whole period & '14-'23 & post-COVID & GFT period & '14-'15 & '15-'16 & '16-'17 & '17-'18 & '18-'19 & '19-'20 & '20-'21 & '21-'22 & '22-'23 \\ 
  \hline \hline \multicolumn{1}{l}{RMSE}\\ARGO-C & \textbf{0.276} & \textbf{0.391} & \textbf{0.542} & \textbf{0.327} & \textbf{0.352} & 0.335 & \textbf{0.274} & \textbf{0.248} & \textbf{0.293} & \textbf{0.828} & 0.127 & 0.287 & \textbf{1.136} \\ 
  ARGOX & 0.278 & 0.398 & 0.556 & 0.328 & 0.353 & \textbf{0.332} & 0.280 & 0.251 & 0.296 & 0.857 & 0.117 & 0.317 & 1.142 \\ 
  VAR1 & 3.012 & 2.731 & 2.112 & 2.174 & 2.376 & 1.319 & 2.017 & 5.069 & 3.685 & 5.899 & 0.617 & 0.840 & 2.727 \\ 
  GFT & -- & -- & -- & 0.744 & 0.813 & -- & -- & -- & -- & -- & -- & -- & -- \\ 
  naive & 0.296 & 0.428 & 0.602 & 0.354 & 0.380 & 0.375 & 0.297 & 0.253 & 0.297 & 0.865 & \textbf{0.114} & \textbf{0.255} & 1.354 \\ 
   \hline \hline \multicolumn{1}{l}{MAE}\\
\hline
\end{tabular}
\endgroup
\caption{Comparison of different methods for state-level \%ILI estimation in Maine.  The MSE, MAE, and correlation are reported. The method with the best performance is highlighted in boldface for each metric in each period. \label{tab:state US.ME}} 
\end{table*}
\begin{table*}
\centering
\begingroup\scriptsize
\begin{tabular}{c|lllllllllllll}
  \hline
  & Whole period & '14-'23 & post-COVID & GFT period & '14-'15 & '15-'16 & '16-'17 & '17-'18 & '18-'19 & '19-'20 & '20-'21 & '21-'22 & '22-'23 \\ 
  \hline \hline \multicolumn{1}{l}{RMSE}\\ARGO-C & \textbf{0.479} & \textbf{0.462} & \textbf{0.431} & \textbf{0.540} & \textbf{0.514} & \textbf{0.568} & \textbf{0.593} & 0.446 & 0.511 & 0.646 & 0.135 & \textbf{0.428} & \textbf{0.791} \\ 
  ARGOX & 0.494 & 0.476 & 0.440 & 0.558 & 0.522 & 0.580 & 0.621 & \textbf{0.438} & \textbf{0.510} & \textbf{0.639} & \textbf{0.126} & 0.459 & 0.893 \\ 
  VAR1 & 1.055 & 1.168 & 1.352 & 1.089 & 1.098 & 1.237 & 1.329 & 0.993 & 0.775 & 1.705 & 0.417 & 0.806 & 3.443 \\ 
  GFT & -- & -- & -- & 1.282 & 1.387 & -- & -- & -- & -- & -- & -- & -- & -- \\ 
  naive & 0.601 & 0.565 & 0.491 & 0.714 & 0.743 & 0.657 & 0.721 & 0.619 & 0.624 & 0.810 & 0.134 & 0.564 & 0.893 \\ 
   \hline \hline \multicolumn{1}{l}{MAE}\\
\hline
\end{tabular}
\endgroup
\caption{Comparison of different methods for state-level \%ILI estimation in Maryland.  The MSE, MAE, and correlation are reported. The method with the best performance is highlighted in boldface for each metric in each period. \label{tab:state US.MD}} 
\end{table*}
\begin{table*}
\centering
\begingroup\scriptsize
\begin{tabular}{c|lllllllllllll}
  \hline
  & Whole period & '14-'23 & post-COVID & GFT period & '14-'15 & '15-'16 & '16-'17 & '17-'18 & '18-'19 & '19-'20 & '20-'21 & '21-'22 & '22-'23 \\ 
  \hline \hline \multicolumn{1}{l}{RMSE}\\ARGO-C & \textbf{0.242} & \textbf{0.326} & \textbf{0.442} & 0.275 & \textbf{0.296} & 0.230 & \textbf{0.190} & \textbf{0.369} & \textbf{0.188} & 0.621 & 0.115 & \textbf{0.426} & \textbf{0.897} \\ 
  ARGOX & 0.249 & 0.331 & 0.446 & \textbf{0.275} & 0.297 & \textbf{0.223} & 0.204 & 0.381 & 0.213 & \textbf{0.619} & 0.101 & 0.445 & 0.911 \\ 
  VAR1 & 0.720 & 0.864 & 1.080 & 0.648 & 0.621 & 0.816 & 0.892 & 0.419 & 0.652 & 1.012 & 0.415 & 0.920 & 2.514 \\ 
  GFT & -- & -- & -- & 0.344 & 0.353 & -- & -- & -- & -- & -- & -- & -- & -- \\ 
  naive & 0.329 & 0.415 & 0.540 & 0.335 & 0.363 & 0.258 & 0.271 & 0.541 & 0.324 & 0.711 & \textbf{0.085} & 0.500 & 1.229 \\ 
   \hline \hline \multicolumn{1}{l}{MAE}\\
\hline
\end{tabular}
\endgroup
\caption{Comparison of different methods for state-level \%ILI estimation in Massachusetts.  The MSE, MAE, and correlation are reported. The method with the best performance is highlighted in boldface for each metric in each period. \label{tab:state US.MA}} 
\end{table*}
\begin{table*}
\centering
\begingroup\scriptsize
\begin{tabular}{c|lllllllllllll}
  \hline
  & Whole period & '14-'23 & post-COVID & GFT period & '14-'15 & '15-'16 & '16-'17 & '17-'18 & '18-'19 & '19-'20 & '20-'21 & '21-'22 & '22-'23 \\ 
  \hline \hline \multicolumn{1}{l}{RMSE}\\ARGO-C & 0.276 & \textbf{0.250} & \textbf{0.194} & \textbf{0.366} & \textbf{0.398} & 0.207 & 0.332 & 0.338 & 0.298 & \textbf{0.316} & 0.132 & \textbf{0.223} & 0.329 \\ 
  ARGOX & \textbf{0.275} & 0.252 & 0.201 & 0.384 & 0.419 & \textbf{0.199} & \textbf{0.328} & \textbf{0.319} & \textbf{0.285} & 0.344 & \textbf{0.130} & 0.238 & \textbf{0.326} \\ 
  VAR1 & 1.020 & 1.029 & 1.046 & 0.980 & 1.048 & 0.646 & 1.471 & 2.000 & 0.605 & 1.188 & 0.173 & 0.408 & 2.644 \\ 
  GFT & -- & -- & -- & 1.203 & 1.302 & -- & -- & -- & -- & -- & -- & -- & -- \\ 
  naive & 0.359 & 0.329 & 0.266 & 0.538 & 0.591 & 0.257 & 0.411 & 0.489 & 0.329 & 0.437 & 0.132 & 0.245 & 0.520 \\ 
   \hline \hline \multicolumn{1}{l}{MAE}\\
\hline
\end{tabular}
\endgroup
\caption{Comparison of different methods for state-level \%ILI estimation in Michigan.  The MSE, MAE, and correlation are reported. The method with the best performance is highlighted in boldface for each metric in each period. \label{tab:state US.MI}} 
\end{table*}
\begin{table*}
\centering
\begingroup\scriptsize
\begin{tabular}{c|lllllllllllll}
  \hline
  & Whole period & '14-'23 & post-COVID & GFT period & '14-'15 & '15-'16 & '16-'17 & '17-'18 & '18-'19 & '19-'20 & '20-'21 & '21-'22 & '22-'23 \\ 
  \hline \hline \multicolumn{1}{l}{RMSE}\\ARGO-C & \textbf{0.571} & \textbf{0.568} & \textbf{0.561} & \textbf{0.611} & \textbf{0.658} & \textbf{0.560} & \textbf{0.415} & 0.407 & \textbf{0.503} & \textbf{1.187} & 0.239 & \textbf{0.963} & \textbf{0.666} \\ 
  ARGOX & 0.578 & 0.593 & 0.619 & 0.653 & 0.706 & 0.565 & 0.420 & \textbf{0.375} & 0.505 & 1.248 & 0.253 & 1.016 & 0.768 \\ 
  VAR1 & 1.153 & 1.164 & 1.185 & 1.557 & 1.693 & 0.758 & 0.812 & 0.934 & 1.185 & 2.129 & 0.198 & 2.103 & 1.641 \\ 
  GFT & -- & -- & -- & 0.780 & 0.850 & -- & -- & -- & -- & -- & -- & -- & -- \\ 
  naive & 0.721 & 0.671 & 0.567 & 0.898 & 0.970 & 0.714 & 0.526 & 0.502 & 0.538 & 1.437 & \textbf{0.092} & 0.984 & 0.770 \\ 
   \hline \hline \multicolumn{1}{l}{MAE}\\
\hline
\end{tabular}
\endgroup
\caption{Comparison of different methods for state-level \%ILI estimation in Minnesota.  The MSE, MAE, and correlation are reported. The method with the best performance is highlighted in boldface for each metric in each period. \label{tab:state US.MN}} 
\end{table*}
\begin{table*}
\centering
\begingroup\scriptsize
\begin{tabular}{c|lllllllllllll}
  \hline
  & Whole period & '14-'23 & post-COVID & GFT period & '14-'15 & '15-'16 & '16-'17 & '17-'18 & '18-'19 & '19-'20 & '20-'21 & '21-'22 & '22-'23 \\ 
  \hline \hline \multicolumn{1}{l}{RMSE}\\ARGO-C & \textbf{0.568} & \textbf{0.561} & \textbf{0.547} & \textbf{0.652} & \textbf{0.713} & \textbf{0.353} & \textbf{0.486} & \textbf{0.939} & \textbf{0.632} & \textbf{0.691} & \textbf{0.400} & \textbf{0.428} & \textbf{1.131} \\ 
  ARGOX & 0.610 & 0.604 & 0.591 & 0.755 & 0.828 & 0.354 & 0.488 & 1.027 & 0.636 & 0.763 & 0.400 & 0.438 & 1.270 \\ 
  VAR1 & 2.273 & 2.162 & 1.938 & 2.985 & 3.274 & 1.781 & 1.877 & 4.023 & 2.191 & 2.034 & 1.610 & 1.553 & 3.783 \\ 
  GFT & -- & -- & -- & 1.454 & 1.561 & -- & -- & -- & -- & -- & -- & -- & -- \\ 
  naive & 0.790 & 0.751 & 0.673 & 1.022 & 1.123 & 0.448 & 0.650 & 1.320 & 0.840 & 0.946 & 0.446 & 0.634 & 1.384 \\ 
   \hline \hline \multicolumn{1}{l}{MAE}\\
\hline
\end{tabular}
\endgroup
\caption{Comparison of different methods for state-level \%ILI estimation in Mississippi.  The MSE, MAE, and correlation are reported. The method with the best performance is highlighted in boldface for each metric in each period. \label{tab:state US.MS}} 
\end{table*}
\begin{table*}
\centering
\begingroup\scriptsize
\begin{tabular}{c|lllllllllllll}
  \hline
  & Whole period & '14-'23 & post-COVID & GFT period & '14-'15 & '15-'16 & '16-'17 & '17-'18 & '18-'19 & '19-'20 & '20-'21 & '21-'22 & '22-'23 \\ 
  \hline \hline \multicolumn{1}{l}{RMSE}\\ARGO-C & \textbf{0.658} & \textbf{0.584} & \textbf{0.410} & \textbf{0.531} & \textbf{0.564} & 0.370 & \textbf{0.624} & 0.949 & \textbf{0.807} & \textbf{1.223} & 0.164 & 0.372 & \textbf{0.553} \\ 
  ARGOX & 0.670 & 0.608 & 0.472 & 0.561 & 0.600 & \textbf{0.360} & 0.676 & \textbf{0.946} & 0.840 & 1.335 & 0.154 & \textbf{0.331} & 0.580 \\ 
  VAR1 & 0.876 & 0.759 & 0.470 & 0.895 & 0.971 & 0.394 & 0.761 & 1.378 & 1.012 & 1.542 & \textbf{0.107} & 0.396 & 0.741 \\ 
  GFT & -- & -- & -- & 0.728 & 0.786 & -- & -- & -- & -- & -- & -- & -- & -- \\ 
  naive & 0.843 & 0.744 & 0.511 & 0.911 & 0.989 & 0.393 & 0.692 & 1.368 & 1.008 & 1.474 & 0.112 & 0.406 & 0.726 \\ 
   \hline \hline \multicolumn{1}{l}{MAE}\\
\hline
\end{tabular}
\endgroup
\caption{Comparison of different methods for state-level \%ILI estimation in Missouri.  The MSE, MAE, and correlation are reported. The method with the best performance is highlighted in boldface for each metric in each period. \label{tab:state US.MO}} 
\end{table*}
\begin{table*}
\centering
\begingroup\scriptsize
\begin{tabular}{c|lllllllllllll}
  \hline
  & Whole period & '14-'23 & post-COVID & GFT period & '14-'15 & '15-'16 & '16-'17 & '17-'18 & '18-'19 & '19-'20 & '20-'21 & '21-'22 & '22-'23 \\ 
  \hline \hline \multicolumn{1}{l}{RMSE}\\ARGO-C & \textbf{0.227} & \textbf{0.284} & \textbf{0.368} & \textbf{0.249} & \textbf{0.271} & 0.233 & \textbf{0.141} & 0.157 & 0.439 & \textbf{0.390} & \textbf{0.166} & \textbf{0.349} & \textbf{0.811} \\ 
  ARGOX & 0.229 & 0.291 & 0.381 & 0.254 & 0.277 & \textbf{0.233} & 0.143 & \textbf{0.157} & 0.440 & 0.398 & 0.167 & 0.362 & 0.850 \\ 
  VAR1 & 3.395 & 2.943 & 1.829 & 4.147 & 4.503 & 2.903 & 3.537 & 6.030 & 2.562 & 2.865 & 1.351 & 1.787 & 3.343 \\ 
  GFT & -- & -- & -- & 1.100 & 1.192 & -- & -- & -- & -- & -- & -- & -- & -- \\ 
  naive & 0.241 & 0.317 & 0.424 & 0.296 & 0.323 & 0.259 & 0.154 & 0.174 & \textbf{0.429} & 0.457 & 0.166 & 0.394 & 0.953 \\ 
   \hline \hline \multicolumn{1}{l}{MAE}\\
\hline
\end{tabular}
\endgroup
\caption{Comparison of different methods for state-level \%ILI estimation in Montana.  The MSE, MAE, and correlation are reported. The method with the best performance is highlighted in boldface for each metric in each period. \label{tab:state US.MT}} 
\end{table*}
\begin{table*}
\centering
\begingroup\scriptsize
\begin{tabular}{c|lllllllllllll}
  \hline
  & Whole period & '14-'23 & post-COVID & GFT period & '14-'15 & '15-'16 & '16-'17 & '17-'18 & '18-'19 & '19-'20 & '20-'21 & '21-'22 & '22-'23 \\ 
  \hline \hline \multicolumn{1}{l}{RMSE}\\ARGO-C & \textbf{0.605} & \textbf{0.600} & \textbf{0.591} & \textbf{0.698} & \textbf{0.766} & 0.326 & \textbf{0.466} & 0.823 & \textbf{0.828} & \textbf{0.789} & 0.239 & \textbf{0.362} & \textbf{1.489} \\ 
  ARGOX & 0.617 & 0.617 & 0.617 & 0.707 & 0.777 & 0.303 & 0.492 & \textbf{0.817} & 0.894 & 0.850 & 0.230 & 0.389 & 1.536 \\ 
  VAR1 & 2.160 & 1.991 & 1.632 & 1.568 & 1.711 & 0.491 & 1.314 & 3.915 & 3.141 & 2.887 & 0.415 & 1.269 & 4.152 \\ 
  GFT & -- & -- & -- & 0.866 & 0.951 & -- & -- & -- & -- & -- & -- & -- & -- \\ 
  naive & 0.704 & 0.702 & 0.697 & 0.835 & 0.918 & \textbf{0.286} & 0.521 & 0.913 & 0.972 & 1.100 & \textbf{0.193} & 0.479 & 1.744 \\ 
   \hline \hline \multicolumn{1}{l}{MAE}\\
\hline
\end{tabular}
\endgroup
\caption{Comparison of different methods for state-level \%ILI estimation in Nebraska.  The MSE, MAE, and correlation are reported. The method with the best performance is highlighted in boldface for each metric in each period. \label{tab:state US.NE}} 
\end{table*}
\begin{table*}
\centering
\begingroup\scriptsize
\begin{tabular}{c|lllllllllllll}
  \hline
  & Whole period & '14-'23 & post-COVID & GFT period & '14-'15 & '15-'16 & '16-'17 & '17-'18 & '18-'19 & '19-'20 & '20-'21 & '21-'22 & '22-'23 \\ 
  \hline \hline \multicolumn{1}{l}{RMSE}\\ARGO-C & \textbf{0.308} & \textbf{0.333} & \textbf{0.374} & \textbf{0.365} & \textbf{0.400} & 0.332 & \textbf{0.245} & \textbf{0.215} & \textbf{0.368} & 0.518 & 0.403 & \textbf{0.319} & \textbf{0.370} \\ 
  ARGOX & 0.310 & 0.338 & 0.384 & 0.371 & 0.406 & \textbf{0.332} & 0.254 & 0.221 & 0.369 & 0.527 & \textbf{0.400} & 0.321 & 0.430 \\ 
  VAR1 & 1.351 & 1.313 & 1.240 & 0.557 & 0.580 & 1.095 & 2.185 & 2.248 & 1.059 & 1.985 & 0.479 & 0.887 & 2.850 \\ 
  GFT & -- & -- & -- & 1.658 & 1.705 & -- & -- & -- & -- & -- & -- & -- & -- \\ 
  naive & 0.330 & 0.359 & 0.408 & 0.447 & 0.489 & 0.363 & 0.245 & 0.354 & 0.429 & \textbf{0.394} & 0.418 & 0.354 & 0.550 \\ 
   \hline \hline \multicolumn{1}{l}{MAE}\\
\hline
\end{tabular}
\endgroup
\caption{Comparison of different methods for state-level \%ILI estimation in Nevada.  The MSE, MAE, and correlation are reported. The method with the best performance is highlighted in boldface for each metric in each period. \label{tab:state US.NV}} 
\end{table*}
\begin{table*}
\centering
\begingroup\scriptsize
\begin{tabular}{c|lllllllllllll}
  \hline
  & Whole period & '14-'23 & post-COVID & GFT period & '14-'15 & '15-'16 & '16-'17 & '17-'18 & '18-'19 & '19-'20 & '20-'21 & '21-'22 & '22-'23 \\ 
  \hline \hline \multicolumn{1}{l}{RMSE}\\ARGO-C & 0.407 & \textbf{0.407} & 0.406 & 0.285 & 0.312 & 0.152 & 0.340 & \textbf{0.719} & 0.605 & 0.662 & \textbf{0.230} & 0.584 & \textbf{0.269} \\ 
  ARGOX & \textbf{0.406} & 0.411 & 0.421 & 0.272 & 0.298 & \textbf{0.148} & \textbf{0.327} & 0.727 & \textbf{0.595} & 0.709 & 0.236 & \textbf{0.581} & 0.284 \\ 
  VAR1 & 0.872 & 1.390 & 2.027 & 1.228 & 1.350 & 0.375 & 0.604 & 1.318 & 0.810 & 1.884 & 1.259 & 2.589 & 3.477 \\ 
  GFT & -- & -- & -- & 1.106 & 1.196 & -- & -- & -- & -- & -- & -- & -- & -- \\ 
  naive & 0.436 & 0.419 & \textbf{0.385} & \textbf{0.241} & \textbf{0.264} & 0.165 & 0.346 & 0.797 & 0.612 & \textbf{0.620} & 0.251 & 0.598 & 0.366 \\ 
   \hline \hline \multicolumn{1}{l}{MAE}\\
\hline
\end{tabular}
\endgroup
\caption{Comparison of different methods for state-level \%ILI estimation in New Hampshire.  The MSE, MAE, and correlation are reported. The method with the best performance is highlighted in boldface for each metric in each period. \label{tab:state US.NH}} 
\end{table*}
\begin{table*}
\centering
\begingroup\scriptsize
\begin{tabular}{c|lllllllllllll}
  \hline
  & Whole period & '14-'23 & post-COVID & GFT period & '14-'15 & '15-'16 & '16-'17 & '17-'18 & '18-'19 & '19-'20 & '20-'21 & '21-'22 & '22-'23 \\ 
  \hline \hline \multicolumn{1}{l}{RMSE}\\ARGO-C & \textbf{0.446} & \textbf{0.623} & \textbf{0.859} & \textbf{0.470} & \textbf{0.511} & \textbf{0.485} & 0.620 & \textbf{0.716} & \textbf{0.196} & \textbf{1.575} & 0.266 & \textbf{0.810} & \textbf{0.773} \\ 
  ARGOX & 0.461 & 0.638 & 0.876 & 0.498 & 0.541 & 0.485 & 0.676 & 0.718 & 0.238 & 1.590 & 0.252 & 0.848 & 0.812 \\ 
  VAR1 & 1.548 & 1.842 & 2.290 & 1.157 & 1.221 & 1.580 & 2.953 & 1.932 & 1.062 & 4.069 & 1.344 & 1.738 & 2.497 \\ 
  GFT & -- & -- & -- & 0.828 & 0.860 & -- & -- & -- & -- & -- & -- & -- & -- \\ 
  naive & 0.566 & 0.704 & 0.905 & 0.593 & 0.641 & 0.503 & \textbf{0.605} & 1.004 & 0.431 & 1.625 & \textbf{0.234} & 0.973 & 0.947 \\ 
   \hline \hline \multicolumn{1}{l}{MAE}\\
\hline
\end{tabular}
\endgroup
\caption{Comparison of different methods for state-level \%ILI estimation in New Jersey.  The MSE, MAE, and correlation are reported. The method with the best performance is highlighted in boldface for each metric in each period. \label{tab:state US.NJ}} 
\end{table*}
\begin{table*}
\centering
\begingroup\scriptsize
\begin{tabular}{c|lllllllllllll}
  \hline
  & Whole period & '14-'23 & post-COVID & GFT period & '14-'15 & '15-'16 & '16-'17 & '17-'18 & '18-'19 & '19-'20 & '20-'21 & '21-'22 & '22-'23 \\ 
  \hline \hline \multicolumn{1}{l}{RMSE}\\ARGO-C & 0.502 & \textbf{0.573} & \textbf{0.685} & \textbf{0.430} & \textbf{0.466} & \textbf{0.449} & 0.362 & 0.731 & 0.586 & \textbf{0.856} & \textbf{0.248} & 0.535 & \textbf{1.687} \\ 
  ARGOX & \textbf{0.498} & 0.574 & 0.694 & 0.449 & 0.488 & 0.465 & \textbf{0.347} & \textbf{0.689} & \textbf{0.567} & 0.928 & 0.249 & \textbf{0.498} & 1.691 \\ 
  VAR1 & 2.238 & 2.942 & 3.927 & 2.220 & 2.402 & 2.092 & 1.849 & 2.396 & 2.605 & 4.151 & 1.208 & 3.901 & 9.287 \\ 
  GFT & -- & -- & -- & 1.456 & 1.510 & -- & -- & -- & -- & -- & -- & -- & -- \\ 
  naive & 0.647 & 0.726 & 0.852 & 0.665 & 0.729 & 0.484 & 0.409 & 0.911 & 0.749 & 1.136 & 0.256 & 0.706 & 2.174 \\ 
   \hline \hline \multicolumn{1}{l}{MAE}\\
\hline
\end{tabular}
\endgroup
\caption{Comparison of different methods for state-level \%ILI estimation in New Mexico.  The MSE, MAE, and correlation are reported. The method with the best performance is highlighted in boldface for each metric in each period. \label{tab:state US.NM}} 
\end{table*}
\begin{table*}
\centering
\begingroup\scriptsize
\begin{tabular}{c|lllllllllllll}
  \hline
  & Whole period & '14-'23 & post-COVID & GFT period & '14-'15 & '15-'16 & '16-'17 & '17-'18 & '18-'19 & '19-'20 & '20-'21 & '21-'22 & '22-'23 \\ 
  \hline \hline \multicolumn{1}{l}{RMSE}\\ARGO-C & \textbf{0.585} & \textbf{0.575} & 0.555 & \textbf{0.665} & \textbf{0.702} & 0.523 & \textbf{0.990} & 0.801 & 0.321 & 0.871 & 0.179 & 0.444 & \textbf{1.103} \\ 
  ARGOX & 0.589 & 0.576 & \textbf{0.551} & 0.690 & 0.734 & \textbf{0.520} & 1.064 & \textbf{0.719} & \textbf{0.294} & \textbf{0.839} & 0.167 & 0.447 & 1.131 \\ 
  VAR1 & 1.260 & 1.593 & 2.074 & 1.217 & 1.310 & 0.886 & 1.402 & 1.417 & 1.764 & 3.741 & 1.135 & 1.597 & 1.939 \\ 
  GFT & -- & -- & -- & 0.860 & 0.916 & -- & -- & -- & -- & -- & -- & -- & -- \\ 
  naive & 0.688 & 0.669 & 0.632 & 0.838 & 0.899 & 0.580 & 1.060 & 0.992 & 0.379 & 1.044 & \textbf{0.150} & \textbf{0.430} & 1.384 \\ 
   \hline \hline \multicolumn{1}{l}{MAE}\\
\hline
\end{tabular}
\endgroup
\caption{Comparison of different methods for state-level \%ILI estimation in New York.  The MSE, MAE, and correlation are reported. The method with the best performance is highlighted in boldface for each metric in each period. \label{tab:state US.NY}} 
\end{table*}
\begin{table*}
\centering
\begingroup\scriptsize
\begin{tabular}{c|lllllllllllll}
  \hline
  & Whole period & '14-'23 & post-COVID & GFT period & '14-'15 & '15-'16 & '16-'17 & '17-'18 & '18-'19 & '19-'20 & '20-'21 & '21-'22 & '22-'23 \\ 
  \hline \hline \multicolumn{1}{l}{RMSE}\\ARGO-C & \textbf{0.460} & \textbf{0.428} & \textbf{0.363} & \textbf{0.537} & \textbf{0.589} & \textbf{0.344} & \textbf{0.711} & \textbf{0.716} & 0.335 & 0.372 & 0.121 & \textbf{0.202} & \textbf{0.941} \\ 
  ARGOX & 0.492 & 0.458 & 0.387 & 0.696 & 0.765 & 0.359 & 0.726 & 0.739 & \textbf{0.328} & \textbf{0.352} & 0.112 & 0.205 & 1.012 \\ 
  VAR1 & 1.197 & 1.910 & 2.785 & 1.171 & 1.250 & 0.474 & 1.204 & 0.990 & 1.307 & 2.437 & 0.686 & 2.770 & 6.903 \\ 
  GFT & -- & -- & -- & 0.644 & 0.704 & -- & -- & -- & -- & -- & -- & -- & -- \\ 
  naive & 0.693 & 0.617 & 0.442 & 1.022 & 1.125 & 0.456 & 0.917 & 1.065 & 0.541 & 0.634 & \textbf{0.093} & 0.324 & 1.078 \\ 
   \hline \hline \multicolumn{1}{l}{MAE}\\
\hline
\end{tabular}
\endgroup
\caption{Comparison of different methods for state-level \%ILI estimation in North Carolina.  The MSE, MAE, and correlation are reported. The method with the best performance is highlighted in boldface for each metric in each period. \label{tab:state US.NC}} 
\end{table*}
\begin{table*}
\centering
\begingroup\scriptsize
\begin{tabular}{c|lllllllllllll}
  \hline
  & Whole period & '14-'23 & post-COVID & GFT period & '14-'15 & '15-'16 & '16-'17 & '17-'18 & '18-'19 & '19-'20 & '20-'21 & '21-'22 & '22-'23 \\ 
  \hline \hline \multicolumn{1}{l}{RMSE}\\ARGO-C & \textbf{0.767} & \textbf{0.889} & \textbf{1.079} & \textbf{0.899} & \textbf{0.990} & \textbf{0.432} & \textbf{1.099} & 0.776 & \textbf{0.610} & \textbf{1.284} & \textbf{0.394} & \textbf{0.999} & \textbf{2.578} \\ 
  ARGOX & 0.786 & 0.911 & 1.106 & 0.902 & 0.992 & 0.435 & 1.114 & 0.796 & 0.648 & 1.335 & 0.396 & 1.061 & 2.631 \\ 
  VAR1 & 1.158 & 1.427 & 1.824 & 1.061 & 1.161 & 0.618 & 1.377 & \textbf{0.729} & 0.885 & 2.886 & 0.855 & 1.935 & 3.853 \\ 
  GFT & -- & -- & -- & 1.020 & 1.112 & -- & -- & -- & -- & -- & -- & -- & -- \\ 
  naive & 0.891 & 1.011 & 1.202 & 0.960 & 1.056 & 0.495 & 1.342 & 0.988 & 0.775 & 1.480 & 0.412 & 1.077 & 2.861 \\ 
   \hline \hline \multicolumn{1}{l}{MAE}\\
\hline
\end{tabular}
\endgroup
\caption{Comparison of different methods for state-level \%ILI estimation in North Dakota.  The MSE, MAE, and correlation are reported. The method with the best performance is highlighted in boldface for each metric in each period. \label{tab:state US.ND}} 
\end{table*}
\begin{table*}
\centering
\begingroup\scriptsize
\begin{tabular}{c|lllllllllllll}
  \hline
  & Whole period & '14-'23 & post-COVID & GFT period & '14-'15 & '15-'16 & '16-'17 & '17-'18 & '18-'19 & '19-'20 & '20-'21 & '21-'22 & '22-'23 \\ 
  \hline \hline \multicolumn{1}{l}{RMSE}\\ARGO-C & \textbf{0.301} & \textbf{0.343} & \textbf{0.410} & \textbf{0.517} & \textbf{0.568} & 0.222 & \textbf{0.235} & \textbf{0.347} & 0.317 & \textbf{0.416} & 0.187 & \textbf{0.407} & \textbf{0.888} \\ 
  ARGOX & 0.321 & 0.374 & 0.455 & 0.580 & 0.636 & \textbf{0.214} & 0.239 & 0.352 & \textbf{0.315} & 0.478 & \textbf{0.187} & 0.436 & 1.015 \\ 
  VAR1 & 1.528 & 1.458 & 1.320 & 1.907 & 2.055 & 0.733 & 2.348 & 2.549 & 0.952 & 2.274 & 0.248 & 0.911 & 2.513 \\ 
  GFT & -- & -- & -- & 0.954 & 1.039 & -- & -- & -- & -- & -- & -- & -- & -- \\ 
  naive & 0.426 & 0.473 & 0.549 & 0.783 & 0.861 & 0.255 & 0.322 & 0.526 & 0.379 & 0.515 & 0.189 & 0.515 & 1.334 \\ 
   \hline \hline \multicolumn{1}{l}{MAE}\\
\hline
\end{tabular}
\endgroup
\caption{Comparison of different methods for state-level \%ILI estimation in Ohio.  The MSE, MAE, and correlation are reported. The method with the best performance is highlighted in boldface for each metric in each period. \label{tab:state US.OH}} 
\end{table*}
\begin{table*}
\centering
\begingroup\scriptsize
\begin{tabular}{c|lllllllllllll}
  \hline
  & Whole period & '14-'23 & post-COVID & GFT period & '14-'15 & '15-'16 & '16-'17 & '17-'18 & '18-'19 & '19-'20 & '20-'21 & '21-'22 & '22-'23 \\ 
  \hline \hline \multicolumn{1}{l}{RMSE}\\ARGO-C & 0.955 & 0.901 & \textbf{0.790} & 1.444 & 1.576 & 0.890 & 1.552 & 0.853 & \textbf{0.725} & \textbf{1.018} & 0.579 & 1.014 & \textbf{1.099} \\ 
  ARGOX & \textbf{0.933} & \textbf{0.889} & 0.801 & 1.435 & 1.564 & 0.887 & 1.500 & \textbf{0.733} & 0.747 & 1.068 & \textbf{0.577} & \textbf{0.996} & 1.121 \\ 
  VAR1 & 3.239 & 2.751 & 1.465 & 4.118 & 4.533 & 2.011 & 5.221 & 3.008 & 3.988 & 4.228 & 0.960 & 1.142 & 1.758 \\ 
  GFT & -- & -- & -- & 2.056 & 2.208 & -- & -- & -- & -- & -- & -- & -- & -- \\ 
  naive & 0.987 & 0.925 & 0.796 & \textbf{1.375} & \textbf{1.508} & \textbf{0.887} & \textbf{1.457} & 1.015 & 0.994 & 1.176 & 0.618 & 1.002 & 1.130 \\ 
   \hline \hline \multicolumn{1}{l}{MAE}\\
\hline
\end{tabular}
\endgroup
\caption{Comparison of different methods for state-level \%ILI estimation in Oklahoma.  The MSE, MAE, and correlation are reported. The method with the best performance is highlighted in boldface for each metric in each period. \label{tab:state US.OK}} 
\end{table*}
\begin{table*}
\centering
\begingroup\scriptsize
\begin{tabular}{c|lllllllllllll}
  \hline
  & Whole period & '14-'23 & post-COVID & GFT period & '14-'15 & '15-'16 & '16-'17 & '17-'18 & '18-'19 & '19-'20 & '20-'21 & '21-'22 & '22-'23 \\ 
  \hline \hline \multicolumn{1}{l}{RMSE}\\ARGO-C & 0.428 & 0.397 & 0.332 & 0.371 & 0.402 & \textbf{0.569} & 0.817 & 0.339 & \textbf{0.358} & 0.413 & 0.158 & 0.222 & \textbf{0.775} \\ 
  ARGOX & \textbf{0.425} & \textbf{0.393} & \textbf{0.324} & 0.368 & 0.399 & 0.579 & \textbf{0.810} & \textbf{0.290} & 0.367 & \textbf{0.387} & 0.144 & \textbf{0.197} & 0.804 \\ 
  VAR1 & 2.425 & 2.061 & 1.102 & 3.854 & 4.239 & 1.258 & 3.884 & 2.004 & 2.512 & 2.624 & 0.986 & 1.147 & 1.567 \\ 
  GFT & -- & -- & -- & 0.570 & 0.487 & -- & -- & -- & -- & -- & -- & -- & -- \\ 
  naive & 0.500 & 0.478 & 0.433 & \textbf{0.323} & \textbf{0.348} & 0.632 & 0.946 & 0.442 & 0.450 & 0.757 & \textbf{0.127} & 0.283 & 0.891 \\ 
   \hline \hline \multicolumn{1}{l}{MAE}\\
\hline
\end{tabular}
\endgroup
\caption{Comparison of different methods for state-level \%ILI estimation in Oregon.  The MSE, MAE, and correlation are reported. The method with the best performance is highlighted in boldface for each metric in each period. \label{tab:state US.OR}} 
\end{table*}
\begin{table*}
\centering
\begingroup\scriptsize
%
\endgroup
\caption{Comparison of different methods for state-level \%ILI estimation in Pennsylvania.  The MSE, MAE, and correlation are reported. The method with the best performance is highlighted in boldface for each metric in each period. \label{tab:state US.PA}} 
\end{table*}
\begin{table*}
\centering
\begingroup\scriptsize
%
\endgroup
\caption{Comparison of different methods for state-level \%ILI estimation in Rhode Island.  The MSE, MAE, and correlation are reported. The method with the best performance is highlighted in boldface for each metric in each period. \label{tab:state US.RI}} 
\end{table*}
\begin{table*}
\centering
\begingroup\scriptsize
%
\endgroup
\caption{Comparison of different methods for state-level \%ILI estimation in South Carolina.  The MSE, MAE, and correlation are reported. The method with the best performance is highlighted in boldface for each metric in each period. \label{tab:state US.SC}} 
\end{table*}
\begin{table*}
\centering
\begingroup\scriptsize
%
\endgroup
\caption{Comparison of different methods for state-level \%ILI estimation in South Dakota.  The MSE, MAE, and correlation are reported. The method with the best performance is highlighted in boldface for each metric in each period. \label{tab:state US.SD}} 
\end{table*}
\begin{table*}
\centering
\begingroup\scriptsize
%
\endgroup
\caption{Comparison of different methods for state-level \%ILI estimation in Tennessee.  The MSE, MAE, and correlation are reported. The method with the best performance is highlighted in boldface for each metric in each period. \label{tab:state US.TN}} 
\end{table*}
\begin{table*}
\centering
\begingroup\scriptsize
%
\endgroup
\caption{Comparison of different methods for state-level \%ILI estimation in Texas.  The MSE, MAE, and correlation are reported. The method with the best performance is highlighted in boldface for each metric in each period. \label{tab:state US.TX}} 
\end{table*}
\begin{table*}
\centering
\begingroup\scriptsize
%
\endgroup
\caption{Comparison of different methods for state-level \%ILI estimation in New York NY.  The MSE, MAE, and correlation are reported. The method with the best performance is highlighted in boldface for each metric in each period. \label{tab:state US.NYC}} 
\end{table*}

%% file: main.bbl
\begin{thebibliography}{10}
\urlstyle{rm}
\expandafter\ifx\csname url\endcsname\relax
  \def\url#1{\texttt{#1}}\fi
\expandafter\ifx\csname urlprefix\endcsname\relax\def\urlprefix{URL }\fi
\expandafter\ifx\csname doiprefix\endcsname\relax\def\doiprefix{DOI: }\fi
\providecommand{\bibinfo}[2]{#2}
\providecommand{\eprint}[2][]{\url{#2}}

\bibitem{polgreen2008using}
\bibinfo{author}{Polgreen, P.~M.}, \bibinfo{author}{Chen, Y.},
  \bibinfo{author}{Pennock, D.~M.}, \bibinfo{author}{Nelson, F.~D.} \&
  \bibinfo{author}{Weinstein, R.~A.}
\newblock \bibinfo{journal}{\bibinfo{title}{Using internet searches for
  influenza surveillance}}.
\newblock {\emph{\JournalTitle{Clinical infectious diseases}}}
  \textbf{\bibinfo{volume}{47}}, \bibinfo{pages}{1443--1448}
  (\bibinfo{year}{2008}).

\bibitem{ginsberg2009detecting}
\bibinfo{author}{Ginsberg, J.} \emph{et~al.}
\newblock \bibinfo{journal}{\bibinfo{title}{Detecting influenza epidemics using
  search engine query data}}.
\newblock {\emph{\JournalTitle{Nature}}} \textbf{\bibinfo{volume}{457}},
  \bibinfo{pages}{1012--1014} (\bibinfo{year}{2009}).

\bibitem{althouse2011prediction}
\bibinfo{author}{Althouse, B.~M.}, \bibinfo{author}{Ng, Y.~Y.} \&
  \bibinfo{author}{Cummings, D.~A.}
\newblock \bibinfo{journal}{\bibinfo{title}{Prediction of dengue incidence
  using search query surveillance}}.
\newblock {\emph{\JournalTitle{PLoS neglected tropical diseases}}}
  \textbf{\bibinfo{volume}{5}}, \bibinfo{pages}{e1258} (\bibinfo{year}{2011}).

\bibitem{chan2011using}
\bibinfo{author}{Chan, E.~H.}, \bibinfo{author}{Sahai, V.},
  \bibinfo{author}{Conrad, C.} \& \bibinfo{author}{Brownstein, J.~S.}
\newblock \bibinfo{journal}{\bibinfo{title}{Using web search query data to
  monitor dengue epidemics: a new model for neglected tropical disease
  surveillance}}.
\newblock {\emph{\JournalTitle{PLoS Neglected Tropical Diseases}}}
  \textbf{\bibinfo{volume}{5}}, \bibinfo{pages}{e1206} (\bibinfo{year}{2011}).

\bibitem{murdoch2013inevitable}
\bibinfo{author}{Murdoch, T.~B.} \& \bibinfo{author}{Detsky, A.~S.}
\newblock \bibinfo{journal}{\bibinfo{title}{The inevitable application of big
  data to health care}}.
\newblock {\emph{\JournalTitle{Jama}}} \textbf{\bibinfo{volume}{309}},
  \bibinfo{pages}{1351--1352} (\bibinfo{year}{2013}).

\bibitem{lee2013real}
\bibinfo{author}{Lee, K.}, \bibinfo{author}{Agrawal, A.} \&
  \bibinfo{author}{Choudhary, A.}
\newblock \bibinfo{title}{Real-time disease surveillance using twitter data:
  demonstration on flu and cancer}.
\newblock In \emph{\bibinfo{booktitle}{Proceedings of the 19th ACM SIGKDD
  international conference on Knowledge discovery and data mining}},
  \bibinfo{pages}{1474--1477} (\bibinfo{year}{2013}).

\bibitem{khoury2014big}
\bibinfo{author}{Khoury, M.~J.} \& \bibinfo{author}{Ioannidis, J.~P.}
\newblock \bibinfo{journal}{\bibinfo{title}{Big data meets public health}}.
\newblock {\emph{\JournalTitle{Science}}} \textbf{\bibinfo{volume}{346}},
  \bibinfo{pages}{1054--1055} (\bibinfo{year}{2014}).

\bibitem{rufai2020world}
\bibinfo{author}{Rufai, S.~R.} \& \bibinfo{author}{Bunce, C.}
\newblock \bibinfo{journal}{\bibinfo{title}{World leaders’ usage of twitter
  in response to the covid-19 pandemic: a content analysis}}.
\newblock {\emph{\JournalTitle{Journal of public health}}}
  \textbf{\bibinfo{volume}{42}}, \bibinfo{pages}{510--516}
  (\bibinfo{year}{2020}).

\bibitem{effenberger2020association}
\bibinfo{author}{Effenberger, M.} \emph{et~al.}
\newblock \bibinfo{journal}{\bibinfo{title}{Association of the covid-19
  pandemic with internet search volumes: a google trendstm analysis}}.
\newblock {\emph{\JournalTitle{International Journal of Infectious Diseases}}}
  \textbf{\bibinfo{volume}{95}}, \bibinfo{pages}{192--197}
  (\bibinfo{year}{2020}).

\bibitem{aiello2020social}
\bibinfo{author}{Aiello, A.~E.}, \bibinfo{author}{Renson, A.} \&
  \bibinfo{author}{Zivich, P.}
\newblock \bibinfo{journal}{\bibinfo{title}{Social media-and internet-based
  disease surveillance for public health}}.
\newblock {\emph{\JournalTitle{Annual review of public health}}}
  \textbf{\bibinfo{volume}{41}}, \bibinfo{pages}{101} (\bibinfo{year}{2020}).

\bibitem{lampos2021tracking}
\bibinfo{author}{Lampos, V.} \emph{et~al.}
\newblock \bibinfo{journal}{\bibinfo{title}{Tracking covid-19 using online
  search}}.
\newblock {\emph{\JournalTitle{NPJ digital medicine}}}
  \textbf{\bibinfo{volume}{4}}, \bibinfo{pages}{17} (\bibinfo{year}{2021}).

\bibitem{ettredge2005using}
\bibinfo{author}{Ettredge, M.}, \bibinfo{author}{Gerdes, J.} \&
  \bibinfo{author}{Karuga, G.}
\newblock \bibinfo{journal}{\bibinfo{title}{Using web-based search data to
  predict macroeconomic statistics}}.
\newblock {\emph{\JournalTitle{Communications of the ACM}}}
  \textbf{\bibinfo{volume}{48}}, \bibinfo{pages}{87--92}
  (\bibinfo{year}{2005}).

\bibitem{goel2010predicting}
\bibinfo{author}{Goel, S.}, \bibinfo{author}{Hofman, J.~M.},
  \bibinfo{author}{Lahaie, S.}, \bibinfo{author}{Pennock, D.~M.} \&
  \bibinfo{author}{Watts, D.~J.}
\newblock \bibinfo{journal}{\bibinfo{title}{Predicting consumer behavior with
  web search}}.
\newblock {\emph{\JournalTitle{Proceedings of the National Academy of
  Sciences}}} \textbf{\bibinfo{volume}{107}}, \bibinfo{pages}{17486--17490}
  (\bibinfo{year}{2010}).

\bibitem{mclaren2011using}
\bibinfo{author}{McLaren, N.} \& \bibinfo{author}{Shanbhogue, R.}
\newblock \bibinfo{journal}{\bibinfo{title}{Using internet search data as
  economic indicators}}.
\newblock {\emph{\JournalTitle{Bank of England Quarterly Bulletin}}}
  \bibinfo{pages}{Q2} (\bibinfo{year}{2011}).

\bibitem{bollen2011twitter}
\bibinfo{author}{Bollen, J.}, \bibinfo{author}{Mao, H.} \&
  \bibinfo{author}{Zeng, X.}
\newblock \bibinfo{journal}{\bibinfo{title}{Twitter mood predicts the stock
  market}}.
\newblock {\emph{\JournalTitle{Journal of computational science}}}
  \textbf{\bibinfo{volume}{2}}, \bibinfo{pages}{1--8} (\bibinfo{year}{2011}).

\bibitem{choi2012predicting}
\bibinfo{author}{Choi, H.} \& \bibinfo{author}{Varian, H.}
\newblock \bibinfo{journal}{\bibinfo{title}{Predicting the present with google
  trends}}.
\newblock {\emph{\JournalTitle{Economic Record}}}
  \textbf{\bibinfo{volume}{88}}, \bibinfo{pages}{2--9} (\bibinfo{year}{2012}).

\bibitem{preis2013quantifying}
\bibinfo{author}{Preis, T.}, \bibinfo{author}{Moat, H.~S.} \&
  \bibinfo{author}{Stanley, H.~E.}
\newblock \bibinfo{journal}{\bibinfo{title}{Quantifying trading behavior in
  financial markets using google trends}}.
\newblock {\emph{\JournalTitle{Scientific reports}}}
  \textbf{\bibinfo{volume}{3}}, \bibinfo{pages}{1--6} (\bibinfo{year}{2013}).

\bibitem{scott2014predicting}
\bibinfo{author}{Scott, S.~L.} \& \bibinfo{author}{Varian, H.~R.}
\newblock \bibinfo{journal}{\bibinfo{title}{Predicting the present with
  bayesian structural time series}}.
\newblock {\emph{\JournalTitle{International Journal of Mathematical Modelling
  and Numerical Optimisation}}} \textbf{\bibinfo{volume}{5}},
  \bibinfo{pages}{4--23} (\bibinfo{year}{2014}).

\bibitem{einav2014economics}
\bibinfo{author}{Einav, L.} \& \bibinfo{author}{Levin, J.}
\newblock \bibinfo{journal}{\bibinfo{title}{Economics in the age of big data}}.
\newblock {\emph{\JournalTitle{Science}}} \textbf{\bibinfo{volume}{346}},
  \bibinfo{pages}{1243089} (\bibinfo{year}{2014}).

\bibitem{wu2015future}
\bibinfo{author}{Wu, L.} \& \bibinfo{author}{Brynjolfsson, E.}
\newblock \bibinfo{title}{The future of prediction: How google searches
  foreshadow housing prices and sales}.
\newblock In \emph{\bibinfo{booktitle}{Economic analysis of the digital
  economy}}, \bibinfo{pages}{89--118} (\bibinfo{publisher}{University of
  Chicago Press}, \bibinfo{year}{2015}).

\bibitem{vicente2015forecasting}
\bibinfo{author}{Vicente, M.~R.}, \bibinfo{author}{L{\'o}pez-Men{\'e}ndez,
  A.~J.} \& \bibinfo{author}{P{\'e}rez, R.}
\newblock \bibinfo{journal}{\bibinfo{title}{Forecasting unemployment with
  internet search data: Does it help to improve predictions when job
  destruction is skyrocketing?}}
\newblock {\emph{\JournalTitle{Technological Forecasting and Social Change}}}
  \textbf{\bibinfo{volume}{92}}, \bibinfo{pages}{132--139}
  (\bibinfo{year}{2015}).

\bibitem{scott2015bayesian}
\bibinfo{author}{Scott, S.~L.} \& \bibinfo{author}{Varian, H.~R.}
\newblock \bibinfo{title}{Bayesian variable selection for nowcasting economic
  time series}.
\newblock In \emph{\bibinfo{booktitle}{Economic analysis of the digital
  economy}}, \bibinfo{pages}{119--135} (\bibinfo{publisher}{University of
  Chicago Press}, \bibinfo{year}{2015}).

\bibitem{yi2021forecasting}
\bibinfo{author}{Yi, D.}, \bibinfo{author}{Ning, S.}, \bibinfo{author}{Chang,
  C.-J.} \& \bibinfo{author}{Kou, S.}
\newblock \bibinfo{journal}{\bibinfo{title}{Forecasting unemployment using
  internet search data via prism}}.
\newblock {\emph{\JournalTitle{Journal of the American Statistical
  Association}}} \textbf{\bibinfo{volume}{116}}, \bibinfo{pages}{1662--1673}
  (\bibinfo{year}{2021}).

\bibitem{manyika2011big}
\bibinfo{author}{Manyika, J.} \emph{et~al.}
\newblock \emph{\bibinfo{title}{Big data: The next frontier for innovation,
  competition, and productivity}} (\bibinfo{publisher}{McKinsey \& Company},
  \bibinfo{year}{2011}).

\bibitem{mcafee2012big}
\bibinfo{author}{McAfee, A.} \& \bibinfo{author}{Brynjolfsson, E.}
\newblock \bibinfo{journal}{\bibinfo{title}{Big data: The management
  revolution}}.
\newblock {\emph{\JournalTitle{Harvard Business Review}}}
  \textbf{\bibinfo{volume}{90}}, \bibinfo{pages}{60--68}
  (\bibinfo{year}{2012}).

\bibitem{chen2012business}
\bibinfo{author}{Chen, H.}, \bibinfo{author}{Chiang, R.~H.} \&
  \bibinfo{author}{Storey, V.~C.}
\newblock \bibinfo{journal}{\bibinfo{title}{Business intelligence and
  analytics: From big data to big impact.}}
\newblock {\emph{\JournalTitle{MIS Quarterly}}} \textbf{\bibinfo{volume}{36}},
  \bibinfo{pages}{1165--1188} (\bibinfo{year}{2012}).

\bibitem{risteski2014can}
\bibinfo{author}{Risteski, D.} \& \bibinfo{author}{Davcev, D.}
\newblock \bibinfo{title}{Can we use daily internet search query data to
  improve predicting power of egarch models for financial time series
  volatility}.
\newblock In \emph{\bibinfo{booktitle}{Proceedings of the International
  Conference on Computer Science and Information Systems (ICSIS'2014), October
  17--18, 2014, Dubai (United Arab Emirates)}} (\bibinfo{year}{2014}).

\bibitem{zhu2019big}
\bibinfo{author}{Zhu, C.}
\newblock \bibinfo{journal}{\bibinfo{title}{Big data as a governance
  mechanism}}.
\newblock {\emph{\JournalTitle{The Review of Financial Studies}}}
  \textbf{\bibinfo{volume}{32}}, \bibinfo{pages}{2021--2061}
  (\bibinfo{year}{2019}).

\bibitem{kim2014big}
\bibinfo{author}{Kim, G.-H.}, \bibinfo{author}{Trimi, S.} \&
  \bibinfo{author}{Chung, J.-H.}
\newblock \bibinfo{journal}{\bibinfo{title}{Big-data applications in the
  government sector}}.
\newblock {\emph{\JournalTitle{Communications of the ACM}}}
  \textbf{\bibinfo{volume}{57}}, \bibinfo{pages}{78--85}
  (\bibinfo{year}{2014}).

\bibitem{bennett2007netflix}
\bibinfo{author}{Bennett, J.} \& \bibinfo{author}{Lanning, S.}
\newblock \bibinfo{title}{The netflix prize}.
\newblock In \emph{\bibinfo{booktitle}{Proceedings of KDD Cup and Workshop
  2007}} (\bibinfo{year}{2007}).

\bibitem{santillana2014can}
\bibinfo{author}{Santillana, M.}, \bibinfo{author}{Zhang, D.~W.},
  \bibinfo{author}{Althouse, B.~M.} \& \bibinfo{author}{Ayers, J.~W.}
\newblock \bibinfo{journal}{\bibinfo{title}{What can digital disease detection
  learn from (an external revision to) google flu trends?}}
\newblock {\emph{\JournalTitle{American journal of preventive medicine}}}
  \textbf{\bibinfo{volume}{47}}, \bibinfo{pages}{341--347}
  (\bibinfo{year}{2014}).

\bibitem{wojcik2014public}
\bibinfo{author}{W{\'o}jcik, O.~P.}, \bibinfo{author}{Brownstein, J.~S.},
  \bibinfo{author}{Chunara, R.} \& \bibinfo{author}{Johansson, M.~A.}
\newblock \bibinfo{journal}{\bibinfo{title}{Public health for the people:
  participatory infectious disease surveillance in the digital age}}.
\newblock {\emph{\JournalTitle{Emerging themes in epidemiology}}}
  \textbf{\bibinfo{volume}{11}}, \bibinfo{pages}{1--7} (\bibinfo{year}{2014}).

\bibitem{bates2017tracking}
\bibinfo{author}{Bates, M.}
\newblock \bibinfo{journal}{\bibinfo{title}{Tracking disease: digital
  epidemiology offers new promise in predicting outbreaks}}.
\newblock {\emph{\JournalTitle{IEEE pulse}}} \textbf{\bibinfo{volume}{8}},
  \bibinfo{pages}{18--22} (\bibinfo{year}{2017}).

\bibitem{li2020retrospective}
\bibinfo{author}{Li, C.} \emph{et~al.}
\newblock \bibinfo{journal}{\bibinfo{title}{Retrospective analysis of the
  possibility of predicting the covid-19 outbreak from internet searches and
  social media data, china, 2020}}.
\newblock {\emph{\JournalTitle{Eurosurveillance}}}
  \textbf{\bibinfo{volume}{25}}, \bibinfo{pages}{2000199}
  (\bibinfo{year}{2020}).

\bibitem{ma2022using}
\bibinfo{author}{Ma, S.}, \bibinfo{author}{Sun, Y.} \& \bibinfo{author}{Yang,
  S.}
\newblock \bibinfo{journal}{\bibinfo{title}{Using internet search data to
  forecast covid-19 trends: A systematic review}}.
\newblock {\emph{\JournalTitle{Analytics}}} \textbf{\bibinfo{volume}{1}},
  \bibinfo{pages}{210--227} (\bibinfo{year}{2022}).

\bibitem{ma2022covid}
\bibinfo{author}{Ma, S.} \& \bibinfo{author}{Yang, S.}
\newblock \bibinfo{journal}{\bibinfo{title}{Covid-19 forecasts using internet
  search information in the united states}}.
\newblock {\emph{\JournalTitle{Scientific Reports}}}
  \textbf{\bibinfo{volume}{12}}, \bibinfo{pages}{11539} (\bibinfo{year}{2022}).

\bibitem{iuliano2018estimates}
\bibinfo{author}{Iuliano, A.~D.} \emph{et~al.}
\newblock \bibinfo{journal}{\bibinfo{title}{Estimates of global seasonal
  influenza-associated respiratory mortality: a modelling study}}.
\newblock {\emph{\JournalTitle{The Lancet}}} \textbf{\bibinfo{volume}{391}},
  \bibinfo{pages}{1285--1300} (\bibinfo{year}{2018}).

\bibitem{molinari2007annual}
\bibinfo{author}{Molinari, N.-A.~M.} \emph{et~al.}
\newblock \bibinfo{journal}{\bibinfo{title}{The annual impact of seasonal
  influenza in the {US}: measuring disease burden and costs}}.
\newblock {\emph{\JournalTitle{Vaccine}}} \textbf{\bibinfo{volume}{25}},
  \bibinfo{pages}{5086--5096} (\bibinfo{year}{2007}).

\bibitem{lipsitch2011improving}
\bibinfo{author}{Lipsitch, M.}, \bibinfo{author}{Finelli, L.},
  \bibinfo{author}{Heffernan, R.~T.}, \bibinfo{author}{Leung, G.~M.} \&
  \bibinfo{author}{Redd; for~the 2009 H1N1 Surveillance~Group, S.~C.}
\newblock \bibinfo{journal}{\bibinfo{title}{Improving the evidence base for
  decision making during a pandemic: the example of 2009 influenza a/h1n1}}.
\newblock {\emph{\JournalTitle{Biosecurity and bioterrorism: biodefense
  strategy, practice, and science}}} \textbf{\bibinfo{volume}{9}},
  \bibinfo{pages}{89--115} (\bibinfo{year}{2011}).

\bibitem{nsoesie2014systematic}
\bibinfo{author}{Nsoesie, E.~O.}, \bibinfo{author}{Brownstein, J.~S.},
  \bibinfo{author}{Ramakrishnan, N.} \& \bibinfo{author}{Marathe, M.~V.}
\newblock \bibinfo{journal}{\bibinfo{title}{A systematic review of studies on
  forecasting the dynamics of influenza outbreaks}}.
\newblock {\emph{\JournalTitle{Influenza and other respiratory viruses}}}
  \textbf{\bibinfo{volume}{8}}, \bibinfo{pages}{309--316}
  (\bibinfo{year}{2014}).

\bibitem{chretien2014influenza}
\bibinfo{author}{Chretien, J.-P.}, \bibinfo{author}{George, D.},
  \bibinfo{author}{Shaman, J.}, \bibinfo{author}{Chitale, R.~A.} \&
  \bibinfo{author}{McKenzie, F.~E.}
\newblock \bibinfo{journal}{\bibinfo{title}{Influenza forecasting in human
  populations: a scoping review}}.
\newblock {\emph{\JournalTitle{PloS One}}} \textbf{\bibinfo{volume}{9}},
  \bibinfo{pages}{e94130} (\bibinfo{year}{2014}).

\bibitem{brownstein2009digital}
\bibinfo{author}{Brownstein, J.~S.}, \bibinfo{author}{Freifeld, C.~C.} \&
  \bibinfo{author}{Madoff, L.~C.}
\newblock \bibinfo{journal}{\bibinfo{title}{Digital disease
  detection—harnessing the web for public health surveillance}}.
\newblock {\emph{\JournalTitle{The New England journal of medicine}}}
  \textbf{\bibinfo{volume}{360}}, \bibinfo{pages}{2153} (\bibinfo{year}{2009}).

\bibitem{dalton2009flutracking}
\bibinfo{author}{Dalton, C.} \emph{et~al.}
\newblock \bibinfo{journal}{\bibinfo{title}{Flutracking: a weekly australian
  community online survey of influenza-like illness in 2006, 2007 and 2008}}.
\newblock {\emph{\JournalTitle{Communicable diseases intelligence quarterly
  report}}} \textbf{\bibinfo{volume}{33}}, \bibinfo{pages}{316--322}
  (\bibinfo{year}{2009}).

\bibitem{achrekar2011predicting}
\bibinfo{author}{Achrekar, H.}, \bibinfo{author}{Gandhe, A.},
  \bibinfo{author}{Lazarus, R.}, \bibinfo{author}{Yu, S.-H.} \&
  \bibinfo{author}{Liu, B.}
\newblock \bibinfo{title}{Predicting flu trends using twitter data}.
\newblock In \emph{\bibinfo{booktitle}{2011 IEEE conference on computer
  communications workshops (INFOCOM WKSHPS)}}, \bibinfo{pages}{702--707}
  (\bibinfo{organization}{IEEE}, \bibinfo{year}{2011}).

\bibitem{yuan2013monitoring}
\bibinfo{author}{Yuan, Q.} \emph{et~al.}
\newblock \bibinfo{journal}{\bibinfo{title}{Monitoring influenza epidemics in
  china with search query from baidu}}.
\newblock {\emph{\JournalTitle{PloS one}}} \textbf{\bibinfo{volume}{8}},
  \bibinfo{pages}{e64323} (\bibinfo{year}{2013}).

\bibitem{paul2014twitter}
\bibinfo{author}{Paul, M.~J.}, \bibinfo{author}{Dredze, M.} \&
  \bibinfo{author}{Broniatowski, D.}
\newblock \bibinfo{journal}{\bibinfo{title}{Twitter improves influenza
  forecasting}}.
\newblock {\emph{\JournalTitle{PLoS currents}}} \textbf{\bibinfo{volume}{6}}
  (\bibinfo{year}{2014}).

\bibitem{mciver2014wikipedia}
\bibinfo{author}{McIver, D.~J.} \& \bibinfo{author}{Brownstein, J.~S.}
\newblock \bibinfo{journal}{\bibinfo{title}{Wikipedia usage estimates
  prevalence of influenza-like illness in the united states in near
  real-time}}.
\newblock {\emph{\JournalTitle{PLoS computational biology}}}
  \textbf{\bibinfo{volume}{10}}, \bibinfo{pages}{e1003581}
  (\bibinfo{year}{2014}).

\bibitem{santillana2014using}
\bibinfo{author}{Santillana, M.}, \bibinfo{author}{Nsoesie, E.~O.},
  \bibinfo{author}{Mekaru, S.~R.}, \bibinfo{author}{Scales, D.} \&
  \bibinfo{author}{Brownstein, J.~S.}
\newblock \bibinfo{journal}{\bibinfo{title}{Using clinicians’ search query
  data to monitor influenza epidemics}}.
\newblock {\emph{\JournalTitle{Clinical Infectious Diseases}}}
  \textbf{\bibinfo{volume}{59}}, \bibinfo{pages}{1446--1450}
  (\bibinfo{year}{2014}).

\bibitem{paolotti2014web}
\bibinfo{author}{Paolotti, D.} \emph{et~al.}
\newblock \bibinfo{journal}{\bibinfo{title}{Web-based participatory
  surveillance of infectious diseases: the influenzanet participatory
  surveillance experience}}.
\newblock {\emph{\JournalTitle{Clinical Microbiology and Infection}}}
  \textbf{\bibinfo{volume}{20}}, \bibinfo{pages}{17--21}
  (\bibinfo{year}{2014}).

\bibitem{smolinski2015flu}
\bibinfo{author}{Smolinski, M.~S.} \emph{et~al.}
\newblock \bibinfo{journal}{\bibinfo{title}{Flu near you: crowdsourced symptom
  reporting spanning 2 influenza seasons}}.
\newblock {\emph{\JournalTitle{American journal of public health}}}
  \textbf{\bibinfo{volume}{105}}, \bibinfo{pages}{2124--2130}
  (\bibinfo{year}{2015}).

\bibitem{santillana2015combining}
\bibinfo{author}{Santillana, M.} \emph{et~al.}
\newblock \bibinfo{journal}{\bibinfo{title}{Combining search, social media, and
  traditional data sources to improve influenza surveillance}}.
\newblock {\emph{\JournalTitle{PLoS Comput Biol}}}
  \textbf{\bibinfo{volume}{11}}, \bibinfo{pages}{e1004513}
  (\bibinfo{year}{2015}).

\bibitem{yang2017using}
\bibinfo{author}{Yang, S.} \emph{et~al.}
\newblock \bibinfo{journal}{\bibinfo{title}{Using electronic health records and
  internet search information for accurate influenza forecasting}}.
\newblock {\emph{\JournalTitle{BMC infectious diseases}}}
  \textbf{\bibinfo{volume}{17}}, \bibinfo{pages}{1--9} (\bibinfo{year}{2017}).

\bibitem{bradshaw2019influenza}
\bibinfo{author}{Bradshaw, B.} \emph{et~al.}
\newblock \bibinfo{journal}{\bibinfo{title}{Influenza surveillance using
  wearable mobile health devices}}.
\newblock {\emph{\JournalTitle{Online Journal of Public Health Informatics}}}
  \textbf{\bibinfo{volume}{11}} (\bibinfo{year}{2019}).

\bibitem{hassan2019social}
\bibinfo{author}{Hassan~Zadeh, A.}, \bibinfo{author}{Zolbanin, H.~M.},
  \bibinfo{author}{Sharda, R.} \& \bibinfo{author}{Delen, D.}
\newblock \bibinfo{journal}{\bibinfo{title}{Social media for nowcasting flu
  activity: Spatio-temporal big data analysis}}.
\newblock {\emph{\JournalTitle{Information Systems Frontiers}}}
  \textbf{\bibinfo{volume}{21}}, \bibinfo{pages}{743--760}
  (\bibinfo{year}{2019}).

\bibitem{viboud2020fitbit}
\bibinfo{author}{Viboud, C.} \& \bibinfo{author}{Santillana, M.}
\newblock \bibinfo{journal}{\bibinfo{title}{Fitbit-informed influenza
  forecasts}}.
\newblock {\emph{\JournalTitle{The Lancet Digital Health}}}
  \textbf{\bibinfo{volume}{2}}, \bibinfo{pages}{e54--e55}
  (\bibinfo{year}{2020}).

\bibitem{cook2011assessing}
\bibinfo{author}{Cook, S.}, \bibinfo{author}{Conrad, C.},
  \bibinfo{author}{Fowlkes, A.~L.} \& \bibinfo{author}{Mohebbi, M.~H.}
\newblock \bibinfo{journal}{\bibinfo{title}{Assessing google flu trends
  performance in the united states during the 2009 influenza virus a (h1n1)
  pandemic}}.
\newblock {\emph{\JournalTitle{PloS one}}} \textbf{\bibinfo{volume}{6}},
  \bibinfo{pages}{e23610} (\bibinfo{year}{2011}).

\bibitem{pervaiz2012flubreaks}
\bibinfo{author}{Pervaiz, F.}, \bibinfo{author}{Pervaiz, M.},
  \bibinfo{author}{Rehman, N.~A.}, \bibinfo{author}{Saif, U.} \emph{et~al.}
\newblock \bibinfo{journal}{\bibinfo{title}{Flubreaks: early epidemic detection
  from google flu trends}}.
\newblock {\emph{\JournalTitle{Journal of medical Internet research}}}
  \textbf{\bibinfo{volume}{14}}, \bibinfo{pages}{e2102} (\bibinfo{year}{2012}).

\bibitem{butler2013google}
\bibinfo{author}{Butler, D.}
\newblock \bibinfo{journal}{\bibinfo{title}{When google got flu wrong: Us
  outbreak foxes a leading web-based method for tracking seasonal flu}}.
\newblock {\emph{\JournalTitle{Nature}}} \textbf{\bibinfo{volume}{494}},
  \bibinfo{pages}{155--157} (\bibinfo{year}{2013}).

\bibitem{olson2013reassessing}
\bibinfo{author}{Olson, D.~R.}, \bibinfo{author}{Konty, K.~J.},
  \bibinfo{author}{Paladini, M.}, \bibinfo{author}{Viboud, C.} \&
  \bibinfo{author}{Simonsen, L.}
\newblock \bibinfo{journal}{\bibinfo{title}{Reassessing google flu trends data
  for detection of seasonal and pandemic influenza: a comparative
  epidemiological study at three geographic scales}}.
\newblock {\emph{\JournalTitle{PLoS computational biology}}}
  \textbf{\bibinfo{volume}{9}}, \bibinfo{pages}{e1003256}
  (\bibinfo{year}{2013}).

\bibitem{lazer2014parable}
\bibinfo{author}{Lazer, D.}, \bibinfo{author}{Kennedy, R.},
  \bibinfo{author}{King, G.} \& \bibinfo{author}{Vespignani, A.}
\newblock \bibinfo{journal}{\bibinfo{title}{The parable of google flu: traps in
  big data analysis}}.
\newblock {\emph{\JournalTitle{science}}} \textbf{\bibinfo{volume}{343}},
  \bibinfo{pages}{1203--1205} (\bibinfo{year}{2014}).

\bibitem{yang2015accurate}
\bibinfo{author}{Yang, S.}, \bibinfo{author}{Santillana, M.} \&
  \bibinfo{author}{Kou, S.~C.}
\newblock \bibinfo{journal}{\bibinfo{title}{Accurate estimation of influenza
  epidemics using google search data via argo}}.
\newblock {\emph{\JournalTitle{Proceedings of the National Academy of
  Sciences}}} \textbf{\bibinfo{volume}{112}}, \bibinfo{pages}{14473--14478}
  (\bibinfo{year}{2015}).

\bibitem{yang2017advances}
\bibinfo{author}{Yang, S.} \emph{et~al.}
\newblock \bibinfo{journal}{\bibinfo{title}{Advances in using internet searches
  to track dengue}}.
\newblock {\emph{\JournalTitle{PLoS computational biology}}}
  \textbf{\bibinfo{volume}{13}}, \bibinfo{pages}{e1005607}
  (\bibinfo{year}{2017}).

\bibitem{ning2019accurate}
\bibinfo{author}{Ning, S.}, \bibinfo{author}{Yang, S.} \& \bibinfo{author}{Kou,
  S.}
\newblock \bibinfo{journal}{\bibinfo{title}{Accurate regional influenza
  epidemics tracking using internet search data}}.
\newblock {\emph{\JournalTitle{Scientific reports}}}
  \textbf{\bibinfo{volume}{9}}, \bibinfo{pages}{5238} (\bibinfo{year}{2019}).

\bibitem{lu2019improved}
\bibinfo{author}{Lu, F.~S.}, \bibinfo{author}{Hattab, M.~W.},
  \bibinfo{author}{Clemente, C.~L.}, \bibinfo{author}{Biggerstaff, M.} \&
  \bibinfo{author}{Santillana, M.}
\newblock \bibinfo{journal}{\bibinfo{title}{Improved state-level influenza
  nowcasting in the united states leveraging internet-based data and network
  approaches}}.
\newblock {\emph{\JournalTitle{Nature communications}}}
  \textbf{\bibinfo{volume}{10}}, \bibinfo{pages}{147} (\bibinfo{year}{2019}).

\bibitem{yang2021use}
\bibinfo{author}{Yang, S.}, \bibinfo{author}{Ning, S.} \& \bibinfo{author}{Kou,
  S.}
\newblock \bibinfo{journal}{\bibinfo{title}{Use internet search data to
  accurately track state level influenza epidemics}}.
\newblock {\emph{\JournalTitle{Scientific reports}}}
  \textbf{\bibinfo{volume}{11}}, \bibinfo{pages}{1--10} (\bibinfo{year}{2021}).

\bibitem{wang2022covid}
\bibinfo{author}{Wang, T.}, \bibinfo{author}{Ma, S.}, \bibinfo{author}{Baek,
  S.} \& \bibinfo{author}{Yang, S.}
\newblock \bibinfo{journal}{\bibinfo{title}{Covid-19 hospitalizations forecasts
  using internet search data}}.
\newblock {\emph{\JournalTitle{Scientific Reports}}}
  \textbf{\bibinfo{volume}{12}}, \bibinfo{pages}{9661} (\bibinfo{year}{2022}).

\bibitem{ma2023joint}
\bibinfo{author}{Ma, S.}, \bibinfo{author}{Ning, S.} \& \bibinfo{author}{Yang,
  S.}
\newblock \bibinfo{journal}{\bibinfo{title}{Joint covid-19 and influenza-like
  illness forecasts in the united states using internet search information}}.
\newblock {\emph{\JournalTitle{Communications Medicine}}}
  \textbf{\bibinfo{volume}{3}}, \bibinfo{pages}{39} (\bibinfo{year}{2023}).

\bibitem{Tibshirani96}
\bibinfo{author}{Tibshirani, R.}
\newblock \bibinfo{journal}{\bibinfo{title}{Regression shrinkage and selection
  via the lasso}}.
\newblock {\emph{\JournalTitle{Journal of the Royal Statistical Society-Series
  B}}} \textbf{\bibinfo{volume}{58}}, \bibinfo{pages}{267--288}
  (\bibinfo{year}{1996}).

\bibitem{Hoerl70}
\bibinfo{author}{Hoerl, A.} \& \bibinfo{author}{Kennard, R.}
\newblock \bibinfo{journal}{\bibinfo{title}{Ridge regression: biased estimation
  for nonorthogonal problems}}.
\newblock {\emph{\JournalTitle{Technometrics}}} \textbf{\bibinfo{volume}{12}},
  \bibinfo{pages}{55--67} (\bibinfo{year}{1970}).

\bibitem{Tibshirani97}
\bibinfo{author}{Tibshirani, R.}
\newblock \bibinfo{journal}{\bibinfo{title}{The lasso method for variable
  selection in the {Cox} model}}.
\newblock {\emph{\JournalTitle{Statistics in Medicine}}}
  \textbf{\bibinfo{volume}{16}}, \bibinfo{pages}{385--395}
  (\bibinfo{year}{1997}).

\bibitem{Lokhorst99}
\bibinfo{author}{Lokhorst, J.}
\newblock \bibinfo{title}{The lasso and generalized linear models}.
\newblock \bibinfo{type}{Tech. Rep.}, \bibinfo{institution}{University of
  Adelaide} (\bibinfo{year}{1999}).

\bibitem{Roth04}
\bibinfo{author}{Roth, V.}
\newblock \bibinfo{journal}{\bibinfo{title}{The generalized lasso}}.
\newblock {\emph{\JournalTitle{IEEE Transactions on Neural Networks}}}
  \textbf{\bibinfo{volume}{15}}, \bibinfo{pages}{16--28}
  (\bibinfo{year}{2004}).

\bibitem{Zou05}
\bibinfo{author}{Zou, H.} \& \bibinfo{author}{Hastie, T.}
\newblock \bibinfo{journal}{\bibinfo{title}{Regularization and variable
  selection via the elastic net}}.
\newblock {\emph{\JournalTitle{Journal of the Royal Statistical Society-Series
  B}}} \textbf{\bibinfo{volume}{67}}, \bibinfo{pages}{301--320}
  (\bibinfo{year}{2005}).

\bibitem{Yuan06}
\bibinfo{author}{Yuan, M.} \& \bibinfo{author}{Lin, Y.}
\newblock \bibinfo{journal}{\bibinfo{title}{Model selection and estimation in
  regression with grouped variables}}.
\newblock {\emph{\JournalTitle{Journal of the Royal Statistical Society-Series
  B}}} \textbf{\bibinfo{volume}{68}}, \bibinfo{pages}{49--67}
  (\bibinfo{year}{2006}).

\bibitem{Meier08}
\bibinfo{author}{Meier, L.}, \bibinfo{author}{van~de Geer, S.} \&
  \bibinfo{author}{Bhlmann, P.}
\newblock \bibinfo{journal}{\bibinfo{title}{The grouped lasso for logistic
  regression}}.
\newblock {\emph{\JournalTitle{Journal of the Royal Statistical Society-Series
  B}}} \textbf{\bibinfo{volume}{70}}, \bibinfo{pages}{53--71}
  (\bibinfo{year}{2008}).

\bibitem{Simon13}
\bibinfo{author}{Simon, N.}, \bibinfo{author}{Friedman, J.},
  \bibinfo{author}{Hastie, T.} \& \bibinfo{author}{Tibshirani, R.}
\newblock \bibinfo{journal}{\bibinfo{title}{A sparse-group lasso}}.
\newblock {\emph{\JournalTitle{Journal of Computational and Graphical
  Statistics}}} \textbf{\bibinfo{volume}{22}}, \bibinfo{pages}{231--245}
  (\bibinfo{year}{2013}).

\bibitem{Ward63}
\bibinfo{author}{Ward, J.~H.}
\newblock \bibinfo{journal}{\bibinfo{title}{Hierarchical grouping to optimize
  an objective function}}.
\newblock {\emph{\JournalTitle{Journal of the American Statistical
  Association}}} \textbf{\bibinfo{volume}{58}}, \bibinfo{pages}{236--244}
  (\bibinfo{year}{1963}).

\bibitem{macqueen1967classification}
\bibinfo{author}{MacQueen, J.}
\newblock \bibinfo{title}{Classification and analysis of multivariate
  observations}.
\newblock In \emph{\bibinfo{booktitle}{5th Berkeley Symp. Math. Statist.
  Probability}}, \bibinfo{pages}{281--297} (\bibinfo{organization}{University
  of California Los Angeles LA USA}, \bibinfo{year}{1967}).

\bibitem{Banerjee11}
\bibinfo{author}{Banerjee, A.} \& \bibinfo{author}{Shan, H.}
\newblock \emph{\bibinfo{title}{Model-Based Clustering. In: Sammut, C., Webb,
  G.I. (eds) Encyclopedia of Machine Learning.}}
  (\bibinfo{publisher}{Springer}, \bibinfo{year}{2011}).

\bibitem{R22}
\bibinfo{author}{{R Core Team}}.
\newblock \emph{\bibinfo{title}{R: A Language and Environment for Statistical
  Computing}}.
\newblock \bibinfo{organization}{R Foundation for Statistical Computing},
  \bibinfo{address}{Vienna, Austria} (\bibinfo{year}{2022}).

\bibitem{van1995python}
\bibinfo{author}{Van~Rossum, G.}, \bibinfo{author}{Drake, F.~L.} \emph{et~al.}
\newblock \emph{\bibinfo{title}{Python reference manual}}
  (\bibinfo{publisher}{Centrum voor Wiskunde en Informatica Amsterdam},
  \bibinfo{year}{1995}).

\bibitem{rousseeuw1987silhouettes}
\bibinfo{author}{Rousseeuw, P.~J.}
\newblock \bibinfo{journal}{\bibinfo{title}{Silhouettes: a graphical aid to the
  interpretation and validation of cluster analysis}}.
\newblock {\emph{\JournalTitle{Journal of computational and applied
  mathematics}}} \textbf{\bibinfo{volume}{20}}, \bibinfo{pages}{53--65}
  (\bibinfo{year}{1987}).

\bibitem{tibshirani2001estimating}
\bibinfo{author}{Tibshirani, R.}, \bibinfo{author}{Walther, G.} \&
  \bibinfo{author}{Hastie, T.}
\newblock \bibinfo{journal}{\bibinfo{title}{Estimating the number of clusters
  in a data set via the gap statistic}}.
\newblock {\emph{\JournalTitle{Journal of the Royal Statistical Society: Series
  B (Statistical Methodology)}}} \textbf{\bibinfo{volume}{63}},
  \bibinfo{pages}{411--423} (\bibinfo{year}{2001}).

\bibitem{CDC_FluView}
\bibinfo{author}{{Center for Disease Control and Preventions}}.
\newblock \bibinfo{title}{Flu activity \& surveillance} (\bibinfo{year}{2023}).
\newblock
  \bibinfo{note}{\url{https://www.cdc.gov/flu/weekly/overview.htm##ILINet},
  Last accessed on 2023-04-13}.

\bibitem{kaufman1990partitioning}
\bibinfo{author}{Kaufman, L.}
\newblock \bibinfo{journal}{\bibinfo{title}{Partitioning around medoids
  (program pam)}}.
\newblock {\emph{\JournalTitle{Finding groups in data}}}
  \textbf{\bibinfo{volume}{344}}, \bibinfo{pages}{68--125}
  (\bibinfo{year}{1990}).

\bibitem{CDC_FluSight}
\bibinfo{author}{{Center for Disease Control and Preventions}}.
\newblock \bibinfo{title}{Flu activity \& surveillance} (\bibinfo{year}{2023}).
\newblock
  \bibinfo{note}{\url{https://www.cdc.gov/flu/weekly/flusight/index.html}, Last
  accessed on 2023-04-13}.

\end{thebibliography}
